\documentclass[11pt]{article}

\pdfoutput=1

\usepackage[]{graphics,graphicx}

\usepackage{pstricks}

\usepackage{blindtext}

\newcommand{\Pd}[2]{\frac{\partial #1}{\partial #2}}
\newcommand{\lb}{\left(}
\newcommand{\rb}{\right)}
\newcommand{\D}{\displaystyle}

\usepackage{setspace} 

\RequirePackage{lineno}

\begin{document} 

 \modulolinenumbers[1]


\begin{center}

{{\LARGE\bf A mechanical model for phase-separation in debris flow}}
\\[10mm]
{Shiva P. Pudasaini
\\[3mm]
\normalsize{Department of Geophysics, Steinmann Institute, University of Bonn}\\
\normalsize{Meckenheimer Allee 176, D-53115, Bonn, Germany}\\[1mm]
\normalsize{E-mail: pudasaini@geo.uni-bonn.de}\\[3mm]
Jan-Thomas Fischer\\[1mm]
Austrian Research Centre for Forests, Department of Natural Hazards\\
Rennweg 1, A-6020, Innsbruck, Austria
\\[10mm]
}
\end{center}
{\bf Abstract:}  Understanding the physics of phase-separation between solid and fluid phases as a mixture mass moves down slope is a long-standing challenge. Here, we propose an extension of the two phase mass flow model (Pudasaini, 2012) by including a new mechanism, called separation-flux, that leads to strong phase-separation in avalanche and debris flows while balancing the enhanced solid flux with the reduced fluid flux. The separation flux mechanism is capable of describing  the dynamically evolving phase-separation and levee formation in a multi-phase, geometrically three-dimensional debris flow. These are often observed phenomena in natural debris flows and industrial processes that involve the transportation of particulate solid-fluid mixture material. The novel separation-flux model includes several dominant physical and mechanical aspects such as pressure gradients, volume fractions of solid and fluid phases and their gradients, shear-rates, flow depth, material friction, viscosity, material densities, topographic constraints, grain size, etc. Due to the inherent separation mechanism, as the mass moves down slope, more and more solid particles are transported to the front and the sides, resulting in solid-rich and mechanically strong frontal surge head, and lateral levees followed by a weaker tail largely consisting of viscous fluid. The primary frontal solid-rich surge head followed by secondary fluid-rich surges is the consequence of phase-separation. Such typical and dominant phase-separation phenomena are revealed for two-phase debris flow simulations. Finally, changes in flow composition, that are explicitly considered by the new modelling approach, result in significant changes of impact pressure estimates. These are highly important in hazard assessment and mitigation planning and highlight the application potential of the new approach.

\section{Introduction}

Phase-separation and levee formation are often observed phenomena in natural two-phase debris flows, pyroclastic flows and granular flows in terrestrial and extra terrestrial environments. Examples include lobe deposits from 1993 Lascar pyroclastic flows, Chile (Felix and Thomas, 2004), debris flow deposit in Spitsbergen, and debris flow levees in Sacagawea (Braat, 2014; de Haas et al., 2015). As common phenomena in mountainous regions, debris flows fundamentally differ from other types of mass flows, e.g., rock avalanches and dilute sediment-laden water floods. Debris flows contain large amounts of water (typically 20-60\% by volume), and thus the mixture composition of solid and fluid governs the rheological properties, and their coupling significantly influences the flow dynamics (Costa, 1988; Iverson, 1997; Pierson, 2005; Pudasaini, 2012). Debris flow runout prediction is a major challenge for hazard mitigation in mountain regions, gullies and valleys. Together with the flow volume and basal topography, the inundation area and runout distance may strongly depend on debris flow composition and its evolution (de Haas et al., 2015) and rheology (Pudasaini, 2012). In the past, several methods have been proposed to predict debris flow dynamics and runout distance including the topography effects (Pudasaini et al., 2005; Rickenmann, 2005; D'Agostino et al., 2010; Fischer et al., 2012; Mergili et al., 2012), flow volume (Takahashi, 1991; Bathurst et al., 1997; Rickenmann, 1999; Crosta et al., 2004; Berti and Simoni, 2007; Pudasaini and Miller, 2013) or material properties and its composition (Iverson and Denlinger, 2001; Pitman and Le, 2005; Pudasaini et al., 2005; Pirulli, 2009; Scheidl and Rickenmann, 2010; Pudasaini, 2012; Pudasaini and Krautblatter, 2014; H\"urlimann et al., 2015; de Haas et al., 2015; Kattel et al., 2016; Kafle et al., 2016).
\\[3mm]
In debris flow mixtures of sediment and fluid down a mountain slope and their deposition in the run-out plain, phase-separation between solid and fluid results in forming solid-dominated lobes, side levee walls and frontal surge heads. These structures are often observed in natural geophysical mass flows (Bartelt and McArdell, 2009; Conway et al., 2010). A typical debris flow, as a two-phase mixture of solid and viscous fluid may result in solid-rich lateral levee and front, and fluid-rich center and back. Advancing fronts of solid-dominated debris flow surges are reported both in natural flows such as Nojiri River debris flow event, in 1987 in Japan, and Jiang Jia Ravine event in 1990 in China, and the large scale debris flow experiments in USGS flume (Iverson, 1997; 2003; Johnson et al., 2012), and also in small scale laboratory experiments (Fairfield, 2011; Battella et al., 2012; Braat, 2014; de Haas et al., 2015).
\\[3mm]
In debris flows, a boulder rich, coarse grained, high intergranular friction front may develop (Iverson, 1997; Major and Iverson, 1999; Iverson and Denlinger, 2001; Pudasaini et al., 2005; McArdell et al., 2007; Schneider et al., 2011). Nevertheless, fine-grained or liquefied debris accumulates behind the frontal material, pushing the frontal boulders forward and aside forming levees (Sharp and Nobles 1953; Iverson, 1997; Major and Iverson, 1999; Iverson et al., 2010; Johnson et al., 2012; Braat, 2014; de Haas et al., 2015). The local size and phase distribution can have a controlling influence on the dynamics of the bulk mixture flow (Gray and Thornton, 2005). The larger particles or the dominating solid-phase at the flow front and margin means that the effective (basal) friction is greater here than in other regions. This may result in lateral instability leading to the formation of lobes and fingers (Pouliquen et al., 1997). Lobes and fingers are also observed in pyroclastic deposits and in the formation of levees in debris flows (Iverson, 1997; Iverson and Vallance, 2001).
\\[3mm]
Solid-rich frontal surges, lobes and channel levees consists of mechanically strong material due to higher effective bulk friction and density in relation to the viscous, muddy mixture of mechanically weaker material in the central-back and tail of the debris body. The proper understanding of mechanically distinctly evolving local composition is very important in accurate description of run-out, inundation areas, and impact forces. These are key in hazard assessment and developing reliable mitigation plans. This is the reason for rapidly increasing scientific interests and the research in understanding and modelling the physics, dynamics and impact of grain segregation, phase-separation, lobe, frontal head and levee formation.
\\[3mm]
Phase-separation (or, segregation) may occur in many industrial and geophysical free-surface flows, i.e., a mixture of granular materials of different grain sizes often exhibits segregation when shaken or transported. In dry granular materials and powder flows, the blending or separation is of considerable practical importance to the process engineering. This includes, the pharmaceutical, mining and food industries that produce and handle huge quantities of granular materials (Gray and Thornton, 2005). 
\\[3mm]
In the case of snow avalanches similar phenomena are observed. Inverse grading, which leads to segregation was found to influence the flow dynamics (Kern et al., 2005). The underlying formation mechanisms, leading to differently sized granules mainly depend on snow temperature (Steinkogler et al., 2015). The size distribution, which is also found in avalanche deposits (Bartelt and McArdell, 2009), vary for different flow types and influence avalanche run out and flow dynamics. However, Bartelt et al. (2012) found that the formation of levees in avalanches may be independent of granule size and that the formation of shear plane depends on the interplay between terrain slope and avalanche mass flux.
\\[3mm]
Experimental debris flow levees and deposition lobes have been reported in Iverson et al. (2010) and Johnson et al. (2012). With laboratory experiments de Haas et al. (2015) investigated effects of debris flow composition on deposition mechanisms, and the resulting runout morphology. They investigated how composition affects debris flow runout and deposition geometry including lobe thickness and width. Their results indicated a clear and optimum relationship between runout and composition of coarse and fine fractions. Size segregation in bi-disperse granular flows is generated in partially filled thin rotating disks (Gray and Hutter, 1997). A small scale laboratory model flow of bi-disperse mixture of solid particles resulted in deposit with large particles in the top free-surface, and small particles in the bottom of the deposit with strong frontal-marginal coarse-ring (Pudasaini and Hutter, 2007). In the experimental debris flows, front surge, phase-separation and levee formation are clearly observed. Some typical phenomena are the surface and burial trajectories of particles, side levees, flow head and central-back channelized flows. Moreover, the excavation reveals gravel-rich lateral levees and liquified muddy interior (Iverson, 1997; 2003; Johnson et al., 2012). Laboratory experiments on debris flows have been conducted in rectangular and transversally curved channels to investigate in detail the phase-separation, levee formation, and solid-rich coarse snout (Fairfield, 2011; Battella et al., 2012; Braat, 2014; de Haas et al., 2015). This reveals a strong phase-separation, frontal surge and levee formation. During the debris motion voids open and close instantaneously. This paves the way for small grains, or fluid, to settle and percolate in these voids causing vertical sorting which results in solid particles or large boulders accumulations at the top of the flow (Gray and Thornton, 2005). Tractionless free-surface and friction at the bottom of the flow leads to vertical velocity differences (Braat, 2014). Due to the higher velocity at the upper part of the flow the free-surface can transport boulder-rich material to the front of the flow. This induces longitudinal sorting. The levees are formed as resistive and coarse-grained debris-flow snouts are shouldered aside by advancing finer-grained debris (Sharp and Nobles, 1953).
\\[3mm]
Based on percolation phenomenon, which states that smaller particles preferentially fall into underlying void space and uplift large particles towards the free-surface, Gray and Thornton (2005) formulated a model for kinetic sieving (Savage and Lun 1988) of large and small particles in avalanches of dry granular materials. Kinetic sieving is a dominant mechanism in dense granular free-surface flows. Based on this idea there has been recent progress in modeling segregation in bi-disperse granular avalanches (Savage and Lun, 1988; Gray and Thornton, 2005; Shearer et al., 2008; Gray and Ancey, 2009; Gray and Kokelaar, 2010). These models are capable of predicting the evolution of the size distribution in laboratory experiments (Golick and Daniels, 2009; Wiederseiner et al., 2011). Furthermore, to explain the formation of lateral levees enriched in coarse grains, Johnson et al. (2012) produced data in large-scale debris-flow experiments and combined it with modeling of particle-size segregation in a vertical plane. By constructing an empirical velocity field and using it with a particle-size segregation model (Gray and Thornton, 2005), Johnson et al. (2012) predicted the segregation and transport of material in the flow in which coarse material segregates to the flow surface, where shear induces the transport to the flow front. Moreover, in the flow head, coarse material is overridden, circulated and finally deposited to form coarse-particle-enriched levees.
\\[3mm]
Although phase-separation between the solid and fluid phases has a phenomenal and profound effect on debris flow dynamics, impacts and deposition morphology, no model and simulation exists to describe and dynamically simulate the generation and propagation of solid-rich lobes, levees, and surges and the muddy viscous and particle-laden dynamical fluid pool in the central-back and the tail, and in general phase-separation, in the flow of debris mixture consisting of sediment particles and viscous fluid.
\\[3mm]
Proper, complete or comprehensive description of these complex flows require a two-phase mixture flow model, including basic flow physics, describing the strong interaction between solid and fluid phases, and at the same time, an enhanced phase-separation mechanism resulting in solid-rich surges and coarse snout, lateral levee formation and liquified pool of highly viscous and muddy body and tail. This is the main task of the present paper. To do so, here, we enhance the general two-phase mass flow model (Pudasaini, 2012) by introducing in it a new separation-flux mechanism to describe the phase-separation phenomena.

\section{Phase-separation Mechanism in Mixture Transport}

\subsection{Two-phase Mass Flow Model}

In two-phase debris mixtures, phases are characterized by different material properties. The fluid phase is characterized by its material density $\rho_{f}$, viscosity $\eta_{f}$  and isotropic stress distribution; whereas the solid phase is characterized by its material density $\rho_{s}$, internal friction angle $\phi$, the basal friction angle $\delta$, an anisotropic stress distribution, and the lateral earth pressure coefficient $K$. The subscripts $s$ and $f$ represent the solid and the fluid phases respectively, with the depth-averaged velocity components for fluid  $\textbf{u}_{f}$ = ($u_{f}$, $v_{f}$) and for solid $\textbf{u}_{s}$ = ($u_{s}$, $v_{s}$) in the down-slope $(x)$ and the cross-slope $(y)$ directions. The total flow depth is denoted by $h$, and the solid-volume fraction by $\alpha_s$ (similarly the fluid volume fraction $\alpha_f = 1- \alpha_s$).
 $\textbf{u}_{f}, \textbf{u}_{s}, h$ and $\alpha_s$ are functions of space and time. The solid and fluid mass balance equations are given by (Pudasaini, 2012):
\begin{eqnarray}
\begin{array}{lll}
\D{\Pd{}{t}{\lb \alpha_s h\rb} + \frac{\partial}{\partial x}{\lb \alpha_s h u_s\rb}
                 + \frac{\partial}{\partial y}{\lb \alpha_s h v_s\rb}=0},
\\[5mm]
\D{\Pd{}{t}{\lb \alpha_f h\rb} + \frac{\partial}{\partial x}{\lb \alpha_f h u_f\rb}
                 + \frac{\partial}{\partial y}{\lb \alpha_f h v_f\rb}=0.}
\end{array}    
\label{Model_Final_Mass}
\end{eqnarray}
 Similarly, momentum conservation equations for the solid and the fluid phases, respectively, are: 
\begin{eqnarray}
\begin{array}{lll}
\resizebox{.935\hsize}{!}{$\D{\Pd{}{t}\biggl [ \alpha_s h \lb u_s - \gamma \mathcal C\lb u_f -u_s \rb \rb \biggr ]
  +\Pd{}{x}\biggl [ \alpha_s h \lb u_s^2 - \gamma \mathcal C\lb u_f^2 -u_s^2 \rb+ \beta_{x_s} \frac{h}{2}\rb \biggr ]
  +\Pd{}{y}\biggl[ \alpha_s h \lb u_sv_s - \gamma \mathcal C\lb u_fv_f -u_sv_s \rb \rb \biggr ]}
\D{=  h\mathcal S_{x_s}}$},\\[5mm]  
\resizebox{.935\hsize}{!}{$\D{\Pd{}{t}\biggl [ \alpha_s h \lb v_s - \gamma \mathcal C\lb v_f -v_s \rb \rb \biggr ]
  +\Pd{}{x}\biggl [ \alpha_s h \lb u_sv_s - \gamma \mathcal C\lb u_fv_f -u_sv_s \rb\rb \biggr ]
  +\Pd{}{y}\left[ \alpha_s h \lb v_s^2 - \gamma \mathcal C\lb v_f^2 -v_s^2\rb+ \beta_{y_s} \frac{h}{2}  \rb \right ]}
\D{=  h\mathcal S_{y_s}}$},\\[5mm]
\resizebox{.935\hsize}{!}{$ \D{\Pd{}{t}\left [ \alpha_f h \lb u_f + \frac{\alpha_s }{\alpha_f}\mathcal C\lb u_f -u_s \rb \rb \right ]
  +\Pd{}{x}\left [ \alpha_f h \lb u_f^2 + \frac{\alpha_s }{\alpha_f}\mathcal C\lb u_f^2 -u_s^2 \rb  + \beta_{x_f} \frac{h}{2}\rb \right ]
  +\Pd{}{y}\left[ \alpha_f h \lb u_fv_f  + \frac{\alpha_s}{\alpha_f}\mathcal C\lb u_fv_f -u_sv_s \rb \rb \right ]
=  h\mathcal S_{x_f}}$},\\[5mm] 
\resizebox{.935\hsize}{!}{$ \D{\Pd{}{t}\left [ \alpha_f h \lb v_f + \frac{\alpha_s }{\alpha_f}\mathcal C\lb v_f -v_s \rb \rb \right ]
  +\Pd{}{x}\left [ \alpha_f h \lb u_fv_f + \frac{\alpha_s }{\alpha_f}\mathcal C\lb u_fv_f -u_sv_s \rb\rb \right ]
  +\Pd{}{y}\left[ \alpha_f h \lb v_f^2  + \frac{\alpha_s}{\alpha_f}\mathcal C\lb v_f^2 -v_s^2 \rb +  \beta_{y_f} \frac{h}{2}\rb \right ]
=  h\mathcal S_{y_f}}$}.
\end{array}    
\label{Model_Final}
\end{eqnarray}
\\[-1mm]
In (\ref{Model_Final}), the source terms are as follows:
\begin{eqnarray}
\mathcal S_{x_s} = \alpha_s\left [g^x - \frac{u_s}{|{\bf u}_s|}\tan\delta p_{b_s} -\varepsilon p_{b_s}\Pd{b}{x}\right ] 
-\varepsilon \alpha_s\gamma p_{b_f}\left [ \Pd{h}{x} + \Pd{b}{x}\right ]
+  C_{DG} \lb u_f - u_s \rb{ |{\bf u}_f - {\bf u}_s|}^{\jmath-1},
\label{Model_Final_ss}\\[2mm]
\hspace{-2.2cm}\mathcal S_{y_s} = \alpha_s\left [ g^y - \frac{v_s}{|{\bf u}_s|}\tan\delta p_{b_s} -\varepsilon p_{b_s}\Pd{b}{y}\right ] 
-\varepsilon \alpha_s\gamma p_{b_f}\left [ \Pd{h}{y} + \Pd{b}{y}\right ]
+ C_{DG} \lb v_f - v_s \rb{ |{\bf u}_f - {\bf u}_s|}^{\jmath-1},
\label{Model_Final_s}
\end{eqnarray}
\vspace{-1mm}
\begin{eqnarray}
\begin{array}{lll}
\D{\mathcal S_{x_f} = \alpha_f\biggl [g^x - \varepsilon   \biggl [\frac{1}{2}p_{b_f}\frac{h}{\alpha_f}\Pd{\alpha_s}{x} +  p_{b_f}\Pd{b}{x}
 -\frac{1}{\alpha_fN_R}\left \{ 2\frac{\partial^2 u_f}{\partial x^2}+  \frac{\partial^2v_f}{\partial y\partial x}
 + \frac{\partial^2 u_f}{\partial y^2} - \frac{\chi u_f}{\varepsilon^2h^2} \right \}} \\[5mm]
 +\D{ \frac{1}{\alpha_fN_{R_\mathcal A}}\left \{ 2\Pd{}{x}\lb \Pd{\alpha_s}{x}\lb u_f - u_s\rb\rb 
 + \Pd{}{y}\lb \Pd{\alpha_s}{x}\lb v_f -v_s\rb + \Pd{\alpha_s}{y}\lb u_f - u_s\rb\rb\right\}
-\frac{\xi\alpha_s\lb u_f -u_s\rb}{\varepsilon^2 \alpha_fN_{R_\mathcal A}h^2} 
 \biggr]\biggl ]}\\[5mm] 
-\D{\frac{1}{\gamma}C_{DG}\lb u_f - u_s \rb{ |{\bf u}_f - {\bf u}_s|}^{\jmath-1}},
\end{array}    
\label{Model_Final_fx}
\end{eqnarray}
\vspace{-3mm}
\begin{eqnarray}
\begin{array}{lll}
\D{\mathcal S_{y_f} = \alpha_f\biggl [g^y - \varepsilon   \biggl [ \frac{1}{2}p_{b_f}\frac{h}{\alpha_f}\Pd{\alpha_s}{y}+  p_{b_f}\Pd{b}{y}
 -\frac{1}{\alpha_fN_R}\left \{ 2\frac{\partial^2 v_f}{\partial y^2}+  \frac{\partial^2u_f}{\partial x\partial y}
 + \frac{\partial^2 v_f}{\partial x^2} - \frac{\chi v_f}{\varepsilon^2h^2} \right \}} \\[5mm]
 + \D{\frac{1}{\alpha_fN_{R_\mathcal A}}\left \{ 2\Pd{}{y}\lb \Pd{\alpha_s}{y}\lb v_f - v_s\rb\rb 
 + \Pd{}{x}\lb \Pd{\alpha_s}{y}\lb u_f -u_s\rb + \Pd{\alpha_s}{x}\lb v_f - v_s\rb\rb\right\}
-\frac{\xi \alpha_s\lb v_f -v_s\rb}{\varepsilon^2 \alpha_fN_{R_\mathcal A}h^2} 
 \biggr]\biggl ]}\\[5mm] 
-\D{\frac{1}{\gamma}C_{DG}\lb v_f - v_s \rb{ |{\bf u}_f - {\bf u}_s|}^{\jmath-1}}.
\end{array}    
\label{Model_Final_fy}
\end{eqnarray}
The pressures and the other parameters involved in the above model equations are as follows:
\begin{eqnarray}
\begin{array}{lll}
\D{\beta_{x_s} = \varepsilon K_x p_{b_s}, \,\,\,\, \beta_{y_s} = \varepsilon K_y p_{b_s},\,\,\,\beta_{x_f} = \beta_{y_f} = \varepsilon p_{b_f},\,\,\, p_{b_f} = - g^z, \,\,\, p_{b_s} = (1-\gamma)p_{b_f},}\\[5mm]
\D{C_{DG} = \frac{\alpha_s \alpha_f(1-\gamma)}{\left [\varepsilon \mathcal U_T\{{\cal P}\mathcal F(Re_p) + (1-{\cal P})\mathcal G(Re_p)\}\right ]^{\jmath}},\,\,\,\,
\mathcal F = \frac{\gamma}{180}\lb\frac{\alpha_f}{\alpha_s} \rb^3 Re_p, \,\,\,\,  \mathcal G= \alpha_f^{M(Re_p) -1},}
\\[5mm]
\D{\gamma =\frac{\rho_f}{\rho_s},\, Re_p = \frac{\rho_f d~ \mathcal U_T}{\eta_f},\, N_R = \frac{\sqrt{gL}H\rho_f}{\alpha_f\eta_f}, N_{R_{\mathcal A}} = \frac{\sqrt{gL}H\rho_f}{\mathcal A \eta_f},\,
\alpha_f = 1-\alpha_s,\, \mathcal A = \mathcal A(\alpha_f).}
\end{array}    
\label{Model_Final_parameters}
\end{eqnarray}
 Equations (\ref{Model_Final_Mass}) are the depth-averaged mass balances for solid and fluid phases respectively, and (\ref{Model_Final}) are the depth-averaged momentum balances for solid (first two equations) and fluid (other two equations) in the $x$- and $y$-directions, respectively.
\\[3mm]
In the above {non-dimensional equations, $x$, $y$ and $z$ are coordinates in} down-slope, cross-slope and flow normal directions, and $g^x$, $g^y$, $g^z$ are the respective components of gravitational acceleration. $L$ and $H$ are the typical length and depth of the flow, $\varepsilon = {H}/{L}$ is the aspect ratio, and $\mu =\tan\delta$ is the basal friction coefficient. $C_{DG}$  is the generalized drag coefficient. Simple linear (laminar-type, at low velocity) or quadratic (turbulent-type, at high velocity) drag is associated with ${\jmath} = 1$ or $2$, respectively. $\mathcal{U}_{T}$ is the terminal velocity of a particle and $\mathcal{P}\in [0,1]$ is a parameter which combines the solid-like ($\mathcal{G}$) and fluid-like ($\mathcal{F}$) drag contributions to flow resistance. $p_{b_{f}}$ and $p_{b_{s}}$ are the effective fluid and solid pressures. $\gamma$ is the density ratio, $\mathcal{C}$ is the virtual mass coefficient (kinetic energy of the fluid phase induced by solid particles), $M$ is a function of the particle Reynolds number ($R_{e_{p}}$), $\chi$ includes vertical shearing of fluid velocity, and $\xi$ takes into account different distributions of $\alpha_s$. $\mathcal{A}$ is the mobility of the fluid at the interface, and $N_{R}$ and $N_{R_{\mathcal{A}}}$ respectively, are the quasi-Reynolds number and mobility-Reynolds number associated with the classical Newtonian and enhanced non-Newtonian fluid viscous stresses. Slope topography is represented by $b = b(x,y)$.

\subsection{A Novel Mechanical Model for Phase-separation in Mixture Flows}

Without any complex mechanical knowledge of the phase-separation mechanism, it can phenomenologically be summarized as follows:  Mass fluxes are described for each material phase in the mass balances (\ref{Model_Final_Mass}), where no explicit phase interaction and separation terms appear. Without any additional separation-fluxes the solid and fluid materials would tend to homogeneously move down-slope, preventing significant to strong phase-separation, which is often observed in field events and laboratory experiments (Iverson, 1997; 2003; Fairfield, 2011; Battella et al., 2012; Johnson et al., 2012; Braat, 2014; de Haas et al., 2015). An increase, e.g., of solid material flux results in more solid material farther downslope, while in the mean time, a respectively reduced separation-flux for the fluid material results in more fluid material farther upslope. The same process can be applied to the lateral direction: solid material fluxes are enhanced outward from the longitudinal central line, and conversely, the fluid material fluxes are enhanced inward to the longitudinal central line. This brings more solid material towards the outer and frontal regions and more fluid material towards the center and back of the flow. Longitudinal or lateral phase separation fluxes can be considered separately as uni-directional phase-separations. More complex and complete situations require combining both the longitudinal and lateral separation-fluxes and the resulting phase-separations, as often observed in field events or laboratory debris flow experiments (de Haas et al., 2015).
\\[3mm]
While mass flux induced phase separation appears to be phenomenologically clear, here, we develop a new mechanical model to describe the phase-separation in two-phase debris flows. This mechanical approach appropriately accounts for {phase-separation processes} in geometrically three-dimensional mixture mass flows. 

\subsubsection{The Relative Velocity}

For the purpose of developing separation flux models, {we formulate the total net force balance taking into account the forces, that appear in the source terms in (\ref{Model_Final_ss}) and the pressure gradients, that appear in the flux term of the solid momentum balance in (\ref{Model_Final}). Additionally, the solid shear forces may be considered in the total net force balance. For simplicity, we consider only the linear drag, i.e., ${\jmath} = 1$. Ignoring the inertial and acceleration terms, the relative velocity between solid and fluid, $u_r^* = u_f - u_s$, is expressed dividing the net force balance by the drag coefficient $C_{DG}$. The relative velocity $u_r^*$ is of key importance for the phase-separation mechanism, which is directly linked to the net forces that appear in the system. The down-slope solid momentum equation indicates that the relative velocity $u_r^*$ results from different effective force contributions, including:
\begin{eqnarray}
\begin{array}{lll}
u_r^* = f\lb F_C, F_P, F_B, F_T, F_S, \dots\rb = \frac{1}{C_{DG}} \,\lb F_C + F_P + F_B + F_T + F_S + \dots\rb,
\end{array}    
\label{Phase_Separation_1}
\end{eqnarray}
where $F_C$ is the Coulomb force, $F_P$ is the force due to the lateral hydraulic pressure gradient, $F_B$ is the force associated with buoyancy, $F_T$ emerges from the topographic pressure gradient, and $F_S$ is related to the shear force. For convenience, written in dimensional form, these force components are:
\begin{eqnarray}
\begin{array}{lll}
F_C = \displaystyle{\mu(1-\gamma)\alpha_s \rho_s g\cos\zeta},\,\,
F_P = \displaystyle{\frac{K_x}{2}(1-\gamma) \rho_sg \cos\zeta\lb h \Pd{\alpha_s}{x}\rb 
                                  + K_x (1-\gamma) \rho_sg \cos\zeta\lb\alpha_s \Pd{h}{x}\rb},\\[5mm]
F_B = \displaystyle{~\gamma\rho_s g\cos\zeta\lb \alpha_s \Pd{h}{x}\rb },\,\,
F_T = \displaystyle{~ \rho_s g\cos\zeta\lb \alpha_s \Pd{b}{x}\rb },\,\,
F_S = \displaystyle{\chi_s ~\alpha_s\mu_s\frac{u_s}{h_s^2}}.
\end{array}    
\label{Phase_Separation_2}
\end{eqnarray}
Similar to the hydraulic pressure gradient, the topographic pressure gradient also may contribute to phase-separation. {Another} force which also contributes to the phase-separation is the velocity shearing (shear-rate) in the $xy\mbox{-},$ $xz$- and $yz$-planes. Although shearing is virtually neglected in the depth-averaged models, it may contribute to phase-separation and can be revived. Furthermore, depending on the flow situation, this shearing may be localized in a thin layer, or otherwise may be present through the entire flow depth (Domnik and Pudasaini, 2012; Domnik et al., 2013). This has been included in the above expression through the approximated shear-rate $\dot\gamma_{xz} \approx u_s/h_s^2$ (any other suitable expression can be used), where $\chi_s$ is the shear factor, and $\mu_s$ is the dynamic viscosity (Pudasaini, 2012), resulting the shear-force $F_S$ in (\ref{Phase_Separation_2}). Moreover, the corresponding expression for the relative velocity $v_r^*$ in the lateral $y$-direction can be constructed. However, any other physically important and relevant forces can be included in (\ref{Phase_Separation_1}), and thus in the list (\ref{Phase_Separation_2}). Examples may include cohesion, and electrostatic forces.

\subsubsection{Separation Fluxes}

Here, we develop a new mechanical model for separation fluxes resulting in separation between solid and fluid in a mixture flow. Let $k$ represents one of the $n$ phases in the mixture and $m$ represents the subscript associated with the (bulk) mixture. The mixture density $\rho_m$, the mass fraction $\mathcal{C}_k$ of the phase $k$, the mixture velocity ${\bf u}_m$, and the mixture mass flux $\rho_m {\bf u}_m$ are given by (Manninen et al., 1996; Krasnopolsky et al., 2016)
\begin{eqnarray}
\begin{array}{lll}
\rho_m = \sum{\alpha_k \rho_k},\,\,\,
\mathcal{C}_k = \D{\frac{\alpha_k \rho_k}{\rho_m }},\,\,\,
{\bf u}_m = \D{\frac{1}{\rho_m}\sum{\alpha_k \rho_k {\bf u}_k} = \sum{\mathcal{C}_k {\bf u}_k}},\,\,\,
\rho_m {\bf u}_m = \D{\sum{\alpha_k \rho_k {\bf u}_k}},
\end{array}    
\label{Variable_Defn_1}
\end{eqnarray}
 where $\sum{\alpha_k = 1}$ is the holdup constraint, and ${\bf u}_m = (u_m, v_m, w_m)$ are the mixture velocity components along the downslope $x$, cross-slope $y$ and flow depth direction $z$, respectively. For any phase $k$, the full dimensional phase mass balance equation is given by:
\begin{eqnarray}
\begin{array}{lll}
\D{\Pd{}{t}\lb\alpha_k\rho_k\rb 
+ \Pd{}{x}\lb \alpha_k\rho_k u_k\rb 
+ \Pd{}{y}\lb \alpha_k\rho_k v_k\rb
+ \Pd{}{z}\lb \alpha_k\rho_k w_k\rb
= 0.}
\end{array}    
\label{MassBalance_Solid_Fluid}
\end{eqnarray}
The slip velocity $u_{f_k}$ of the phase $k$ with respect to the velocity of a background (continuum) phase $f$, and the diffusion velocity $u_{{D_k}}$ of the phase $k$, respectively,  are denoted by
\begin{eqnarray}
\begin{array}{lll}
u_{f_k} = u_k - u_f;\,\,\,\,
u_{D_k} = u_k - u_m.
\end{array}    
\label{Variable_Defn_2}
\end{eqnarray}
 Then, the diffusion velocity $u_{D_l} = u_l - u_m$ of a dispersive phase $l$ can be represented by (Manninen et al., 1996)
\begin{eqnarray}
\begin{array}{lll}
u_{D_l} = u_{f_l} - \sum{\mathcal{C}_k u_{f_k}}.
\end{array}    
\label{Diff_Velocity}
\end{eqnarray}
Here, we are mainly concerned with the two-phase solid-fluid mixture flow. For the mixture consisting of a fluid-phase $f$ and a single dispersive (particle, or) solid-phase $s$ (\ref{Diff_Velocity}) reduces to
 \begin{eqnarray}
\begin{array}{lll}
u_{D_s} = u_{f_s} - \mathcal{C}_s u_{f_s} = \lb 1 - \mathcal{C}_s \rb u_{f_s}
= \lb \mathcal{C}_s - 1 \rb {\tilde u}_r^*,
\end{array}    
\label{Diff_Velocity_Solid} 
\end{eqnarray}
where, $\mathcal{C}_s = \alpha_s\rho_s/\rho_m$, and ${\tilde u}_r^* = -u_{f_s}$ is the relative (or, slip) velocity between the solid and the fluid phases that can be obtained from the momentum balance, as shown in Section 2.2.1. Here, ${\tilde u}_r^*$ indicates the full non-depth-averaged velocity field such that  ${u_r^*}$ is its mean (depth-averaged) value. $u_r^*$ should contain all the forces resulting in slip velocity that yields phase separation. For the ease of notation, here, we drop tilde from the corresponding variables until we again introduce the depth-averaged variables while developing the depth-averaged enhanced mass balance equations. The expressions as in (\ref{Phase_Separation_1}) and (\ref{Phase_Separation_2}) can also be derived for the full non-depth averaged descriptions of the relative velocities $\lb u_r^*, v_r^*, w_r^*\rb$ in all three flow directions. 
\\[3mm]
 The diffusion velocity for the fluid-phase, $u_{D_f}$, can be written as:
 \begin{eqnarray}
\begin{array}{lll}
u_{D_f} = u_f - u_m = u_f - \lb u_s -u_{D_s} \rb =  {u_r^*} + u_{D_s}
=  {u_r^*} + \lb \mathcal{C}_s  - 1 \rb {u_r^*} = \mathcal{C}_s {u_r^*}.
\end{array}    
\label{Diff_Velocity_Fluid} 
\end{eqnarray}
Expressions similar to (\ref{Diff_Velocity_Solid} ) and (\ref{Diff_Velocity_Fluid}) can be analogously derived for the $y$- and $z$- directions:
 \begin{eqnarray}
\begin{array}{lll}
v_{D_s} = \lb \mathcal{C}_s - 1 \rb {v_r^*},\,\,
v_{D_f} = \mathcal{C}_s {v_r^*};\,\,\,\,\,\,
w_{D_s} = \lb \mathcal{C}_s  - 1 \rb {w_r^*},\,\,
w_{D_f} = \mathcal{C}_s {w_r^*}.
\end{array}    
\label{Diff_Velocity_SolidFluid_Others} 
\end{eqnarray}
For a mixture consisting of a solid and a fluid phase (\ref{MassBalance_Solid_Fluid}) reduces, respectively, to the solid and fluid phase mass balance equations as:
\begin{eqnarray}
\begin{array}{lll}
\D{\Pd{}{t}\lb\alpha_s\rho_s\rb 
+ \Pd{}{x}\lb \alpha_s\rho_s u_s\rb 
+ \Pd{}{y}\lb \alpha_s\rho_s v_s\rb
+ \Pd{}{z}\lb \alpha_s\rho_s w_s\rb
= 0,}
\end{array}    
\label{MassBalance_Solid}
\end{eqnarray}
\\[-11mm]
\begin{eqnarray}
\begin{array}{lll}
\D{\Pd{}{t}\lb\alpha_f\rho_f\rb 
+ \Pd{}{x}\lb \alpha_f\rho_f u_f\rb 
+ \Pd{}{y}\lb \alpha_f\rho_f v_f\rb
+ \Pd{}{z}\lb \alpha_f\rho_f w_f\rb
= 0.}
\end{array}    
\label{MassBalance_Fluid}
\end{eqnarray}
Summing up (\ref{MassBalance_Solid}) and (\ref{MassBalance_Fluid}) and using (\ref{Variable_Defn_1}) yields the (bulk) mixture mass balance:
 \begin{eqnarray}
\begin{array}{lll}
\D{\Pd{}{t}\lb \rho_m\rb 
+ \Pd{}{x}\lb \rho_m u_m\rb 
+ \Pd{}{y}\lb \rho_m v_m\rb
+ \Pd{}{z}\lb \rho_m w_m\rb
= 0,}
\end{array}    
\label{MassBalance_Bulk_1}
\end{eqnarray}
which can be written as 
\begin{eqnarray}
\begin{array}{lll}
&\D{
\left [
\Pd{}{t}\lb \alpha_s\rho_s\rb 
+ \Pd{}{x}\lb \alpha_s\rho_s u_m\rb 
+ \Pd{}{y}\lb \alpha_s \rho_s v_m\rb
+ \Pd{}{z}\lb \alpha_s\rho_s w_m\rb
\right ] +}\\[5mm] 
&\D{\left [\Pd{}{t}\lb \alpha_f\rho_f\rb 
+ \Pd{}{x}\lb \alpha_f\rho_f u_m\rb 
+ \Pd{}{y}\lb \alpha_f \rho_f v_m\rb
+ \Pd{}{z}\lb \alpha_f\rho_f w_m\rb
\right ]
= 0.
}
\end{array}    
\label{MassBalance_Bulk_2}
\end{eqnarray}
From (\ref{Variable_Defn_2}) with the diffusion velocities, (\ref{MassBalance_Bulk_2}) reduces to
\begin{eqnarray}
\begin{array}{lll}
&\D{
\left [
\Pd{}{t}\lb \alpha_s\rho_s\rb 
+ \Pd{}{x}\lb \alpha_s\rho_s \Big\{u_s - u_{D_s}\Big\}\rb 
+ \Pd{}{y}\lb \alpha_s\rho_s \Big\{v_s - v_{D_s}\Big\}\rb
+ \Pd{}{z}\lb \alpha_s\rho_s \Big\{w_s - w_{D_s}\Big\}\rb
\right ] +}\\[5mm] 
&\D{\left [\Pd{}{t}\lb \alpha_f\rho_f\rb 
+ \Pd{}{x}\lb \alpha_f\rho_f \Big\{u_f \!-\! u_{D_f}\Big\}\rb 
+ \Pd{}{y}\lb \alpha_f\rho_f \Big\{v_f \!-\! v_{D_f}\Big\}\rb
\!+ \Pd{}{z}\lb \alpha_f\rho_f \Big\{w_f \!-\! w_{D_f}\Big\}\rb
\right ]
\!= 0.
}
\end{array}    
\label{MassBalance_Bulk_3}
\end{eqnarray}
Inserting the expressions for $u_{D_s}, u_{D_f};\,\, v_{D_s}, v_{D_f};\,\, w_{D_s}, w_{D_f}$ from (\ref{Diff_Velocity_Solid}), (\ref{Diff_Velocity_Fluid}) and (\ref{Diff_Velocity_SolidFluid_Others} ), this becomes:
\begin{eqnarray}
\begin{array}{lll}
&\D{
\left [
\Pd{}{t}\lb \alpha_s\rho_s\rb 
\!+\! \Pd{}{x}\lb \alpha_s\rho_s \bigg \{u_s \!+\! \frac{\alpha_f\rho_f}{\rho_m}u_r^*\bigg \}\rb 
\!+\! \Pd{}{y}\lb \alpha_s\rho_s \bigg \{v_s \!+\! \frac{\alpha_f\rho_f}{\rho_m}v_r^*\bigg \}\rb
\!+\! \Pd{}{z}\lb \alpha_s\rho_s \bigg \{w_s \!+\! \frac{\alpha_f\rho_f}{\rho_m}w_r^*\bigg \}\rb
\right ] \!+\!}\\[6mm] 
&\D{\left [\Pd{}{t}\lb \alpha_f\rho_f\rb 
\!+\! \Pd{}{x}\lb \!\alpha_f\rho_f \bigg \{u_f \!-\! \frac{\alpha_s\rho_s}{\rho_m}u_r^*\bigg \}\rb 
\!+\! \Pd{}{y}\lb \!\alpha_f\rho_f \bigg \{v_f \!-\! \frac{\alpha_s\rho_s}{\rho_m}v_r^*\bigg \}\rb
\!+\! \Pd{}{z}\lb \!\alpha_f\rho_f \bigg \{w_f \!-\! \frac{\alpha_s\rho_s}{\rho_m}w_r^*\bigg \}\rb
\right ]
\!=\! 0.
}
\end{array}    
\label{MassBalance_Bulk_4}
\end{eqnarray}
This can be split into dynamically coupled and balanced system of mass conservations for solid and fluid phases:
\begin{eqnarray}
\begin{array}{lll}
\D{
\Pd{}{t}\lb \alpha_s\rho_s\rb 
\!+\! \Pd{}{x}\lb \alpha_s\rho_s \bigg \{u_s + \frac{\alpha_f\rho_f}{\rho_m}u_r^*\bigg \}\rb 
\!+\! \Pd{}{y}\lb \alpha_s\rho_s \bigg \{v_s + \frac{\alpha_f\rho_f}{\rho_m}v_r^*\bigg \}\rb
\!+\! \Pd{}{z}\lb \alpha_s\rho_s \bigg \{w_s + \frac{\alpha_f\rho_f}{\rho_m}w_r^*\bigg \}\rb
\!=\! 0,
}
\end{array}    
\label{MassBalance_Coupled_Solid}
\end{eqnarray} 
\\[-10mm]
\begin{eqnarray}
\begin{array}{lll}
\D{\Pd{}{t}\lb \alpha_f\rho_f\rb 
\!+\! \Pd{}{x}\lb \alpha_f\rho_f \bigg \{u_f - \frac{\alpha_s\rho_s}{\rho_m}u_r^*\bigg \}\rb 
\!+\! \Pd{}{y}\lb \alpha_f\rho_f \bigg \{v_f - \frac{\alpha_s\rho_s}{\rho_m}v_r^*\bigg \}\rb
\!+\! \Pd{}{z}\lb \alpha_f\rho_f \bigg \{w_f - \frac{\alpha_s\rho_s}{\rho_m}w_r^*\bigg \}\rb
\!=\! 0.
}
\end{array}    
\label{MassBalance_Coupled_Fluid}
\end{eqnarray}
This splitting is legitimate and appears naturally when viewed (interpreted) in terms of a coupled system, preserving the total mass balance, because adding (\ref{MassBalance_Coupled_Solid}) and (\ref{MassBalance_Coupled_Fluid}) results in the entire system (\ref{MassBalance_Bulk_1}). The advantage of the dynamically coupled system is that without adding or losing any mass the system (\ref{MassBalance_Coupled_Solid}) and (\ref{MassBalance_Coupled_Fluid}) enhances the solid mass flux by the amount $+\lb\alpha_s\alpha_f\rho_s\rho_f/\rho_m\rb u_r^*$, and the fluid mass flux is reduced by the same amount $-\lb\alpha_s\alpha_f\rho_s\rho_f/\rho_m\rb u_r^*$ in the downslope $x$-direction. The same holds in other flow directions, $y$ and $z$, the transversal and flow depth directions. These enhanced and reduced fluxes for the solid and fluid are separating the solid and fluid mass flows resulting in the phase-separation in the mixture. Furthermore, the new separation flux models (\ref{MassBalance_Coupled_Solid}) and (\ref{MassBalance_Coupled_Fluid}) are free of any empirical parameters and are derived without any assumptions. 
\\[3mm]
Since the material phase densities $\rho_s$ and $\rho_f$ are constant, (\ref{MassBalance_Coupled_Solid}) and (\ref{MassBalance_Coupled_Fluid}) can be further simplified:
\begin{eqnarray}
\begin{array}{lll}
\D{
\Pd{\alpha_s}{t} 
+ \Pd{}{x}\lb \alpha_s u_s + \frac{\gamma\,\alpha_s\alpha_f}{\alpha_s + \gamma \alpha_f} u_r^*\rb 
+ \Pd{}{y}\lb \alpha_s v_s + \frac{\gamma\,\alpha_s\alpha_f}{\alpha_s + \gamma \alpha_f} v_r^*\rb 
+ \Pd{}{z}\lb \alpha_s w_s + \frac{\gamma\,\alpha_s\alpha_f}{\alpha_s + \gamma \alpha_f} w_r^*\rb 
= 0,
}
\end{array}    
\label{MassBalance_Coupled_Solid_Rho}
\\[1mm]
\begin{array}{lll}
\D{\Pd{\alpha_f}{t}
+ \Pd{}{x}\lb \alpha_f u_f - \frac{\alpha_s\alpha_f}{\alpha_s + \gamma \alpha_f} u_r^*\rb 
+ \Pd{}{y}\lb \alpha_f v_f - \frac{\alpha_s\alpha_f}{\alpha_s + \gamma \alpha_f} v_r^*\rb
+ \Pd{}{z}\lb \alpha_f w_f - \frac{\alpha_s\alpha_f}{\alpha_s + \gamma \alpha_f} w_r^*\rb
= 0.
}
\end{array}    
\label{MassBalance_Coupled_Fluid_Rho}
\end{eqnarray}
In the derivation of the depth-averaged models, it is legitimate to assume that the velocity scaling in the flow depth direction can be some orders of magnitude smaller than the velocity scaling in the longitudinal and lateral flow directions (Savage and Hutter, 1989). Then, following the standard procedures (Pitman and Le, 2005; Pudasaini, 2012) the mass balance equations (\ref{MassBalance_Coupled_Solid_Rho}) and (\ref{MassBalance_Coupled_Fluid_Rho}) can be depth averaged to yield:
\begin{eqnarray}
\begin{array}{lll}
\D{
\Pd{}{t}\lb \alpha_s h\rb  
+ \Pd{}{x}\lb \alpha_s h u_s + \frac{\gamma\, h\, \alpha_s\alpha_f}{\alpha_s + \gamma \alpha_f} u_r^*\rb 
+ \Pd{}{y}\lb \alpha_s h v_s + \frac{\gamma\, h\, \alpha_s\alpha_f}{\alpha_s + \gamma \alpha_f} v_r^*\rb 
= 0,
}
\end{array}    
\label{MassBalance_Coupled_Solid_Averaged}
\\[1mm]
\begin{array}{lll}
\D{\Pd{}{t}\lb \alpha_f h\rb
\!+\! \Pd{}{x}\lb \alpha_f h u_f -  \frac{h\, \alpha_s\alpha_f}{\alpha_s + \gamma \alpha_f} u_r^*\rb 
\!+\! \Pd{}{y}\lb \alpha_f h v_f -  \frac{h\, \alpha_s\alpha_f}{\alpha_s + \gamma \alpha_f} v_r^*\rb
= 0,}
\end{array}    
\label{MassBalance_Coupled_Fluid_Averaged}
\end{eqnarray}
where all the dynamical variables involved are now the depth-averaged quantities. 
\\[3mm]
For convenience, we introduce the notations for separation fluxes for solid $\lb S^F_{x_s}, S^F_{y_s}\rb $ and fluid $\lb S^F_{x_f}, S^F_{y_f} \rb$ in the down-slope and cross-slope directions,  respectively, 
\begin{eqnarray}
\begin{array}{lll}
 \D{
S^F_{x_s} =  \lambda_s^p u_r^* \alpha_s \alpha_f h,\,\,
S^F_{x_f} =  \lambda_f^p u_r^* \alpha_s \alpha_f h};\,\,\,\,\,\,\,
\D{
S^F_{y_s} =  \lambda_s^p v_r^* \alpha_s \alpha_f h,\,\,
S^F_{y_f} =  \lambda_f^p v_r^* \alpha_s \alpha_f h,}
\end{array}    
\label{Separation_Fluxes_Notations}
\end{eqnarray}
where $\lambda_s^p = \gamma/\lb \alpha_s + \gamma \alpha_f\rb$ and $\lambda_f^p = 1/\lb \alpha_s + \gamma \alpha_f\rb$. The structure in (\ref{Separation_Fluxes_Notations}) indicates, that the separation fluxes $\lb S^F_{x_{s,f}}; S^F_{y_{s,f}}\rb$ vanish for vanishing $\alpha_s$, $\alpha_f$ and $h$.

\subsubsection{Phase Separation Model Equations}

With the separation fluxes (\ref{Separation_Fluxes_Notations}), the  model equations for the enhanced solid and fluid mass balances (\ref{MassBalance_Coupled_Solid_Averaged}) and (\ref{MassBalance_Coupled_Fluid_Averaged}) yield (Pudasaini, 2015):
\begin{eqnarray}
\begin{array}{lll}
\D{
\Pd{}{t}\lb \alpha_s h\rb  
+ \Pd{}{x}\lb \alpha_s h u_s + S^F_{x_s}\rb 
+ \Pd{}{y}\lb \alpha_s h v_s + S^F_{y_s} \rb 
= 0,
}
\end{array}    
\label{Separation_Fluxes_Solid}
\end{eqnarray} 
\\[-11mm]
\begin{eqnarray}
\begin{array}{lll}
\D{\Pd{}{t}\lb \alpha_f h\rb
\!+\! \Pd{}{x}\lb \alpha_f h u_f - S^F_{x_f}\rb
\!+\! \Pd{}{y}\lb \alpha_f h v_f - S^F_{y_f} \rb
\!=\! 0.
}
\end{array}    
\label{Separation_Fluxes_Fluid}
\end{eqnarray}
Equations (\ref{Separation_Fluxes_Solid}) and (\ref{Separation_Fluxes_Fluid}), together with momentum balances (\ref{Model_Final}), constitute a set of novel mechanical model equations (Pudasaini, 2015) capable of describing a three-dimensional time and spatial evolution of complex phase-separation phenomena in two-phase debris, or mass flows. Importantly, by appropriately switching the signs of the separation fluxes, the phase-separations can be reversed. Such a switch can also be applied to gain mixing in two-phase flows and inherently depends on the material (phase) interaction processes and properties. 
 \\[3mm]
It is important to mention that due to the strong phase interactions, considering only the momentum equations and the applied forces therein would not {yield} realistic phase-separation magnitudes. This is a very important and novel finding for phase-separation in mixture flows. So, in order to model the phase-separation in two-phase debris mixture flows, the separation-fluxes $\lb S^F_{x_s}, S^F_{x_f}\rb$ and $\lb S^F_{y_s}, S^F_{y_f}\rb$, as emerged in the derivations (\ref{Separation_Fluxes_Solid}) and (\ref{Separation_Fluxes_Fluid}), must be considered in the down-slope and cross-slope solid and fluid fluxes, in the solid and fluid mass balances, respectively. As mentioned earlier, this does not affect the mass balances for solid and fluid because, this process does not induce additional mass production or loss. 
\\[3mm]
There are several mechanically important aspects in the separation fluxes in (\ref{Separation_Fluxes_Solid}) and (\ref{Separation_Fluxes_Fluid}). These appear mainly in the solid and fluid separation fluxes (\ref{Separation_Fluxes_Notations}). Flux separations are triggered by inter-phase forces, resulting in the non-zero slip velocities $u_r^*$ and $v_r^*$. These separation fluxes are amplified simultaneously by the solid and fluid volume fractions through their product $\alpha_s \alpha_f$. The magnitude is further amplified or controlled by the total (mixture) flow height and the density ratio. Furthermore, the separation flux magnitude is uniformly reduced by the factor $1/\lb \alpha_s + \gamma \alpha_f\rb$ for all the separation flux components $\lb S^F_{x_s}, S^F_{y_s}\rb $ and $\lb S^F_{x_f}, S^F_{y_f} \rb$. Moreover, the separation fluxes can be triggered in one or more flow directions, and depending on the flow dynamics these can be different as indicated by $u_r^*$ and $v_r^*$, which, in general, can be different quantities. The most important characteristic of the separation fluxes is that separation ceases as soon as one of the components (either $\alpha_s$, or $\alpha_f$) in the mixture vanishes. This is a natural condition, because if the presence of the counter (or, complementary) component vanishes nothing needs to be separated. 

\subsubsection{Model Reduction}

 The relative (or, slip) velocities $\lb u_r^*, v_r^*\rb$, that trigger the phase separation mechanisms, have been presented in Section 2.2.1. In simple situations, it may suffice to consider only some particular force components for $u_r^*$ (and $v_r^*$) in (\ref{Phase_Separation_2}). To begin with, we consider {the pressure gradient} $F_P$. For the sake of simplicity, and without loss of generality, we may assume that the gradient of the total debris is dominated by the gradient of the solid volume fraction, and that $\partial \alpha_s/\partial x$  and $\partial \alpha_s/\partial y$ can be approximated or parameterized (say, $\alpha_{s_{p_x}}$ as a function of $x$, or simply a parameter; similarly $\alpha_{s_{p_y}}$). Alternatively, we could also assume that $\lambda_s^p\lb\partial \alpha_s/\partial x\rb \approx $ const. and $\lambda_s^p\lb\partial \alpha_s/\partial y\rb \approx $ const. Then, the separation fluxes are: 
\begin{eqnarray}
\begin{array}{lll}
S^F_{x_s} = S^R_{x_s} \alpha_s\alpha_f h,\,\,
S^F_{y_s} = S^R_{y_s} \alpha_s\alpha_f h;\,\,\,\,
S^F_{x_f} = S^R_{x_f} \alpha_s\alpha_f h,\,\,
S^F_{y_f} = S^R_{y_f} \alpha_s\alpha_f h,
\end{array}    
\label{Phase_Separation_4}
\end{eqnarray}
{where, the separation-rates $S^R_{x_{s,f}}$ and  $S^R_{y_{s,f}}$ are given by:}
\begin{eqnarray}
\begin{array}{lll}
S^R_{x_s} = \displaystyle{\frac{1}{2}~\frac{K_x}{C_{DG}}(1-\gamma) \rho_s g \cos\zeta\lambda_s^p \lb h\Pd{\alpha_s}{x}\rb},\,\,\,\,\,\,\,\,\,\, 
S^R_{y_s} = \displaystyle{\frac{1}{2}~\frac{K_y}{C_{DG}}(1-\gamma) \rho_s g \cos\zeta\lambda_s^p \lb h\Pd{\alpha_s}{y}\rb};\\[6mm]
S^R_{x_f} = \displaystyle{\frac{1}{2}~\frac{K_x}{C_{DG}}(1-\gamma) \rho_s g \cos\zeta\lambda_f^p \lb h\Pd{\alpha_s}{x}\rb},\,\,\,\,\,\,\,\,\,\,
S^R_{y_f} = \displaystyle{\frac{1}{2}~\frac{K_y}{C_{DG}}(1-\gamma) \rho_s g \cos\zeta\lambda_f^p \lb h\Pd{\alpha_s}{y}\rb}.
\label{Phase_Separation_5}
\end{array}    
\end{eqnarray}
The advantage of utilizing the terminology separation-flux is that it can be considered as a general function of the separation-rate, solid and fluid volume fractions and the flow depth. Consequently, the separation velocity emerges from separation-flux as a function of the relative phase velocity, volume fractions of solid or fluid, and the factor $\lambda_{s,f}^p$. {The relations (\ref{Phase_Separation_4}) indicate that structurally $S^R_{x_s}\alpha_f = \lambda_s^p u_r^* \alpha_f =:u^s_e$ and $S^R_{x_f}\alpha_s = \lambda_f^p u_r^* \alpha_s =:u^f_e$ are the effective down-slope separation velocities for solid and fluid phases, respectively.} Thus, $\lambda_{s,f}^p$ are the separation-rate, or the separation-velocity intensity factors. The effective cross-slope separation velocities for solid and fluid phases can be obtained analogously. This indicates that the effective separation velocity for solid varies with the local volume fraction of fluid and the separation-rate for solid. Some important aspects of (\ref{Phase_Separation_5}) in its form are as follows: As the true solid density approaches the true fluid density, the buoyancy reduced normal load of the solid vanishes. Consequently, solid and fluid are close to neutrally buoyant condition so that both phases tend to move together (Pudasaini, 2012). This reduces the phase-separation intensity. Furthermore, as the drag increases, the two phases come closer, resulting in the reduction of the phase-separation.  
\\[3mm]
In general, depending on the complexity of the flow configuration and phase-separation, all {force components $F_i$ that influence the relative velocities $u_r^*$ and $v_r^*$ can contribute to the separation fluxes}. Then, the separation-rates (\ref{Phase_Separation_5}) can be written in general form as:
 \begin{eqnarray}
 \begin{array}{lll}
 S^R_{x_s} = \displaystyle{\lambda^p_s u_r^*},\,\,
 S^R_{y_s} = \displaystyle{\lambda^p_s v_r^*};\,\,\,\,
 S^R_{x_f} = \displaystyle{\lambda^p_f u_r^*},\,\,
 S^R_{y_f} = \displaystyle{\lambda^p_f v_r^*}.
 \end{array}    
 \label{Phase_Separation_6}
 \end{eqnarray}
Furthermore, in the new separation flux approach, terrain slope changes the apparent forces, and changes of mass flux are represented by separation fluxes. Such observations for snow avalanches (Bartelt et al., 2012) are in line with new model approach.
\\[3mm]
 Concentration gradients, that are key parameters in the separation mechanism (\ref{Phase_Separation_5}), may require proper treatments after an optimum or desired separation has been achieved. This may demand for some upper and/or the lower limits of the solid concentration gradients, which could be important for technical and applied problems. {As in chemical phase-separation process (Hillert, 1956; Cahn and  Hilliard, 1958; Vladimirova et al., 1999), this may result in near phase-equilibrium.} The systematic emergence of {$S^R_{x_{s,f}}$ and $S^R_{y_{s,f}}$} in  (\ref{Separation_Fluxes_Solid}) and (\ref{Separation_Fluxes_Fluid}) result in a complex non-linear {anti-diffusion (see, Section 2.2.5)} of solid volume fraction that acts in combination with the other non-linear expressions in fluid momentum equations in (\ref{Model_Final}) and (\ref{Model_Final_fx})-(\ref{Model_Final_fy}) that resulted from non-Newtonian viscous stress emerging from the solid fraction distribution (the terms associated with $N_{R_A}$). This connection between the solid volume fraction gradients balances the dynamics of the concentration gradients in (\ref{Phase_Separation_5}).
\\[3mm]
The new separation flux model can also be compared to the results of the pioneering work by Gray and Thornton (2005), where segregation originates from pressure gradients. In their formulation, simple drag or the Darcy law was used with bulk velocity in a dry granular mixture consisting of particles of different sizes. Down-slope and cross-slope separations were ignored. Bulk velocity is prescribed, and the transport equation for volume fraction was used without the mixture momentum equations. The segregation-rate for a mixture of small and big solid particles as derived by Gray and Thornton (2005) (also see, Johnson et al., 2012) can be realized from (\ref{Phase_Separation_5}) with {$S^R_{x} = (B/c) g \cos\zeta$, where $B$ is a perturbation constant and $c$ is a simple drag coefficient. So, Gray and Thornton (2005), which includes an ad-hoc constraint on the pressure scaling, can be related to the pressure gradients $F_P$.}
\\[3mm]
Here, we have fundamentally advanced in modelling and simulating two-phase, and geometrically three-dimensional phase-separations in a real two-phase debris flow consisting of a mixture of essentially two different materials, solid particles and viscous fluid. In contrast to Gray and Thornton (2005), in our model, $\lb S^R_{x_f} \neq S^R_{x_s}; S^R_{y_f} \neq S^R_{y_s}\rb$ are possible, resulting from different material properties for the solid and fluid constituents. Our new model includes phase-separation in both flow directions. The new mechanical model includes several contributing factors leading to phase-separation. As the model is fully coupled with solid and fluid phase mass and momentum balances no assumption on the bulk and slip velocities is required. The new method now presents a possibility to simulate time and spatial evolution of phase-separation in mixture flows.

\subsubsection{Up-hill Diffusion and Separation-potential}
 
Here, we show that the enhanced solid mass balance (\ref{Separation_Fluxes_Solid}) describes the phase-separation as an advection, and up-hill diffusion process. Similar analysis also holds for fluid. Up-hill diffusion is a typical process in the direction against the concentration gradient resulting in phase separation in mixture (e.g., alloy) (Hillert, 1956; Cahn and  Hilliard, 1958). This anti-diffusion process essentially leads to phase-separation in the flow of a debris mixture. For simplicity, in this Section, we only consider the flow along down-slope. Without loss of generality, we assume that the variation of $h$ with $x$ is negligible. Then, from (\ref{Separation_Fluxes_Solid}) and (\ref{Phase_Separation_5}), we obtain
\begin{eqnarray}
\begin{array}{lll}
\D{
\Pd{\alpha_s}{t}  
+ \Pd{}{x}\lb \alpha_s u_s\rb 
+ \Pd{}{x}\lb D_s^a\, \Pd{\alpha_s}{x} \rb 
= 0,
}
\end{array}    
\label{Uphill_Diffusion}
\end{eqnarray}
where, the coefficient of anti-diffusion, $D_s^a$, is $\displaystyle{D_s^a = \lb {K_x}/{2C_{DG}}\rb (1-\gamma) \rho_s g \cos\zeta\lambda_s^p\alpha_s\alpha_f h}$. Since $D_s^a > 0$, (\ref{Uphill_Diffusion}) governs an advection, and an anti-diffusion for $\alpha_s$, which advects with $u_s$ and diffuses up-hill with  $D_s^a$. Furthermore, in analogy to the chemical-potential in alloy (Hillert, 1956; Cahn and  Hilliard, 1958; Vladimirova et al., 1999), $\alpha_s$ in the diffusion term in (\ref{Uphill_Diffusion}) is called the separation-potential. However, unlike in the chemical-potential, here, the anti-diffusion coefficient $D_s^a$ is a complex function of several physical and mechanical quantities, and the flow variables, including the solid and fluid volume fractions, and the flow height. It clearly shows that the phase separation is triggered by the concentration gradient $\partial \alpha_s/\partial x$, or effectively the non-zero separation-potential, and it is amplified (or, reduced) by the magnitude of $D_s^a$. Furthermore, the phase-separation ceases either, when the gradient $\partial \alpha_s/\partial x$ is close to zero, or when the flow is neutrally buoyant, or if locally one of the phases is negligible. Phase-separation varies inversely with the drag coefficient, which means the  phase-separation process is faster in a mixture with less viscous fluid than the same in the mixture with more viscous fluid.   

\section{Simulation Set-up, Parameters, and Numerical Method}

To assess the capabilities of the new separation-flux model, two-phase flow simulations are performed down an inclined slope with angle $\zeta = 45^\circ$. Initially, the {uniform mixture consists of 50\% solid particles and 50\% viscous fluid, kept in a triangular-wedge $x \in (0, 50)$\,m, $y \in (-35, 35)$\,m that is released instantaneously. The model and material parameter values chosen for the} simulation are: $\phi= 35^\circ$, $\delta = 15^\circ$,  $\rho_{f} =  1,100$ kg m$^{-3}$,\, $\rho_ s = 2,500$\, kg m$^{-3}$,\, $N_{R} = 30,000$,\, $N_{R_\mathcal{A}} = 1,000$,\, $Re_{p} = 1$,\, $\mathcal{U}_{T} = 1.0$ ms$^{-1}$,\, $\mathcal{P} = 0.5$,\, $\jmath = 1$,\, $\chi = 0$,\,  $\xi = 0$,\, $\mathcal{C} = 0.0$. These parameter selections are based on the physics of two-phase mass flows (Pudasaini, 2012, 2014; Pudasaini and Krautblatter, 2014; Kafle et al., 2016; and Kattel et al., 2016). To highlight the process, the separation-rates $( S^R_{s},  S^R_{f})$ can be estimated as follows: for dilatational flows, with the above values of $\phi$ and $\delta$, $K$ is about 0.3 (Pudasaini and Hutter, 2007), for a relatively dense flow $C_{GD}$ is of order unity (Pudasaini, 2012). Considering a typical flow height  of about 2 m and a gentle deviation of $\partial \alpha_s/\partial x$ from a local equilibrium (e.g., 0.005 m$^{-1}$), for a typical solid fraction of 0.65, the separation-rate intensity factors are obtained as $(\lambda_s^p, \lambda_f^p) \approx (0.55, 1.20)$. Then, the separation rates as estimated from (\ref{Phase_Separation_5}) are $( S^R_{s},  S^R_{f}) \approx (5, 10)$ ms$^{-1}$. The separation rate $S^R_{s} = 5$ ms$^{-1}$ is a reasonable estimate for the enhancement of the solid separation flux on inclined surface. With this, the separation velocity for solid is about 1.75\,ms$^{-1}$. This indicates that phase separation is a fairly fast process. Although it depends on the flow configuration and dynamics, it is a reasonable velocity enhancement for the solid as the flow velocity can be on the order of 10\,ms$^{-1}$. Similar analysis holds for the fluid phase.  Since, for the flow considered here, $(S^R_{s},  S^R_{f}) = (5,10)$\,ms$^{-1}$ produce similar results as (\ref{Phase_Separation_5}), we chose these values for simulation. Otherwise, in real and more complex debris flows $S^R_{s}$, and $S^R_{f}$ should be computed from (\ref{Phase_Separation_5}) or (\ref{Phase_Separation_6}).
\\[3mm]
The model equations (\ref{Separation_Fluxes_Solid}) and (\ref{Separation_Fluxes_Fluid}), together with momentum balances (\ref{Model_Final}) are a set of well-structured, non-linear hyperbolic-parabolic partial differential equations in conservative form with complex source terms (Pudasaini, 2012; Kafle et al., 2016). These model equations are used to compute the total debris depth $h$, solid volume fraction $\alpha_{s}$, velocity components for solid $\left(u_{s}, v_{s}\right)$, and for fluid $\left(u_{f}, v_{f}\right)$ in $x$- and $y$-directions, respectively, as functions of space and time. The model equations are solved in conservative variables $\textbf{W}$ = $( h_{s}, h_{f} , m_{x_s}, m_{y_s}, m_{x_f}, m_{y_f})^{t}$, where $ h_{s}=\alpha_{s}h $,  $ h_{f}=\alpha_{f}h$ are the solid and fluid contributions to the debris mixture, or the flow height; and $(m_{x_s}, m_{y_s})  =  (\alpha_s h u_{s}, \alpha_s h v_{s}$), $(m_{x_f}, m_{y_f})  = (\alpha_f h u_{f}, \alpha_f h v_{f})$, are the solid and fluid momenta. This facilitates numerical integration even when shocks are formed in the field variables (Pudasaini, 2014; Kattel et al., 2016). The high-resolution shock-capturing Total Variation Diminishing Non-Oscillatory Central (TVD-NOC) scheme has been implemented (Tai et al., 2002; Pudasaini and Hutter, 2007; Domnik et al., 2013; Pudasaini and Krautblatter, 2014). Advantages of the applied innovative and unified simulation technique for real two-phase debris flows and the corresponding computational strategy have been explained in Kafle et al. (2016) and Kattel et al. (2016).

\section{Simulating Phase-separation in Mixture Flows}

Next, we simulate phase separation between solid and fluid phases as a two-phase debris mass moves down slope. By applying the two-phase debris flow model (Pudasaini, 2012), enhanced with the new separation-flux mechanism the predominant phenomena are revealed in Fig. \ref{Fig_5} and Fig. \ref{Fig_6} for debris flow simulations.

\subsection{Time Evolution of Bi-directional Phase-separation}

Multi-directional phase separations are observed in many mixture flows. Examples include the debris flow experiments by Johnson et al. (2012) and de Haas et al. (2015). In this section the simulation and analysis of the phase separation in  complex flow situation that simultaneously includes phase separations both in the longitudinal and lateral directions is described. The analysis has been presented with the detailed investigations based on the dynamics of the solid-phase, fluid-phase and the total debris mixture. Most importantly, a frontal surge and largely solid-rich mechanically very strong frontal wall and lateral levee walls, followed by relatively weak viscous fluid in the flow body and tail are observed. 

\subsubsection{Bi-directional Solid Phase-separation}

 Bi-directional phase-separation is presented in Fig. \ref{Fig_5}. The figure displays the time evolution of the phase separation for $t = 0, 1, 2, 3, 4, 4.5$\,s. The left panels A show the bi-directional solid phase-separation. This includes the separation of solid from the fluid in both the down-slope and cross slope directions. To obtain such phase-separation, the simulation considers the enhanced solid separation-fluxes in both the flow directions. Similarly, the reduced fluid separation-fluxes are applied in both the flow directions. As soon as the debris mass is released the solid-phase separation takes place {already} at $t = 1$\,s. This process separates solid from the fluid from initially uniformly mixed debris material and pushes the solid material to the sides and to the front of the dominantly downward moving debris body. Although phase separations are effective in both directions, immediately after the flow release, the solid separation appears to be dominating in the lateral direction until $t = 1$\,s, afterwards, separation also becomes stronger in the down-slope direction. Lateral solid walls begin to develop around $t = 1$\,s. Then, these levees are pushed outward and apart in the cross-slope. This process intensifies for $t > 1$\,s. As the down-slope solid separation gains pace for $t \ge 2$\,s, the solid levees are connected in the front of the moving mass, thus forming frontal surge of solid material. The intensity of the lateral separation of solid also depends, including other forces, on the competition between the gravity forces, pressure gradients, and the phase separation-fluxes. As time elapses, from $t = 2$\,s to  $t = 4.5$\,s the separation intensifies. In time, the phase-separation induced strong solid lateral levees and a frontal surge-head are amplified. As the solid mass migrates rapidly to the lateral sides and to the front of the flow, due to the mass balance, the amount of solid in the back and the center of flow is largely reduced. This forms a beautiful three-dimensional double-wing shock wave of solid fraction. This coupled {hydrodynamic} shock-wave is sharpened and consolidated in the front head, and progressively widens and thins in the rear-lateral portion of the solid mass. The emergence, evolution and propagation of such a fish-tail-like solid structure is novel in simulating two-phase solid-fluid mixture flow.   
\begin{figure}[t!]
\vspace{6mm}
\begin{center}
\hbox{\hspace{2.5ex}
 \hspace{-11mm}
 \includegraphics[width=6.9cm, height=2.53cm]{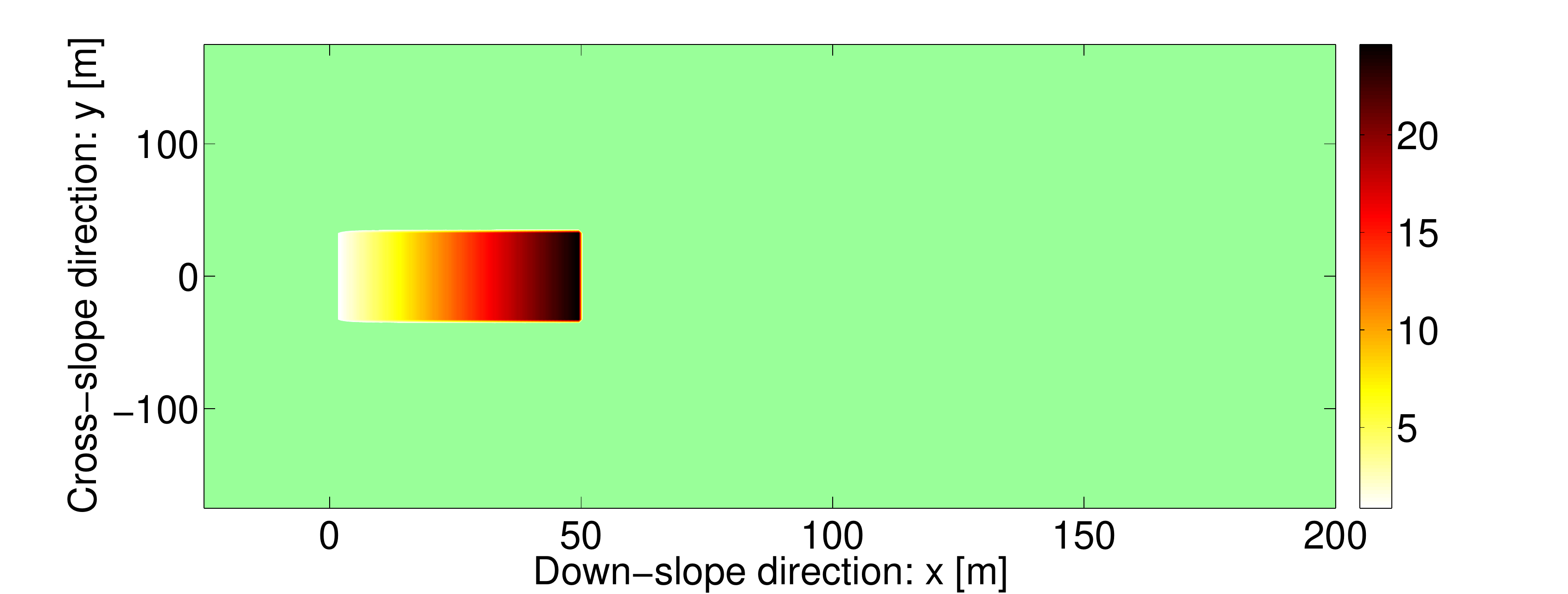}\hspace{-7.mm}
 \includegraphics[width=6.9cm]{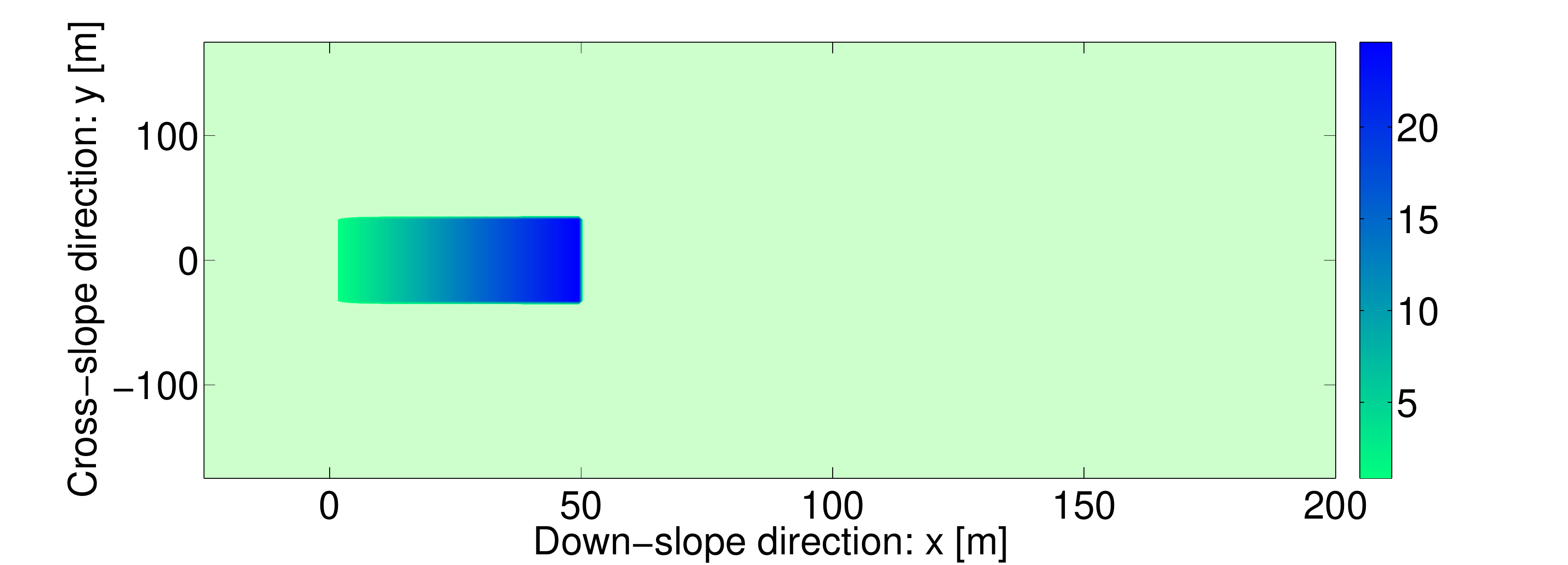}\hspace{-7.mm}
 \includegraphics[width=6.9cm]{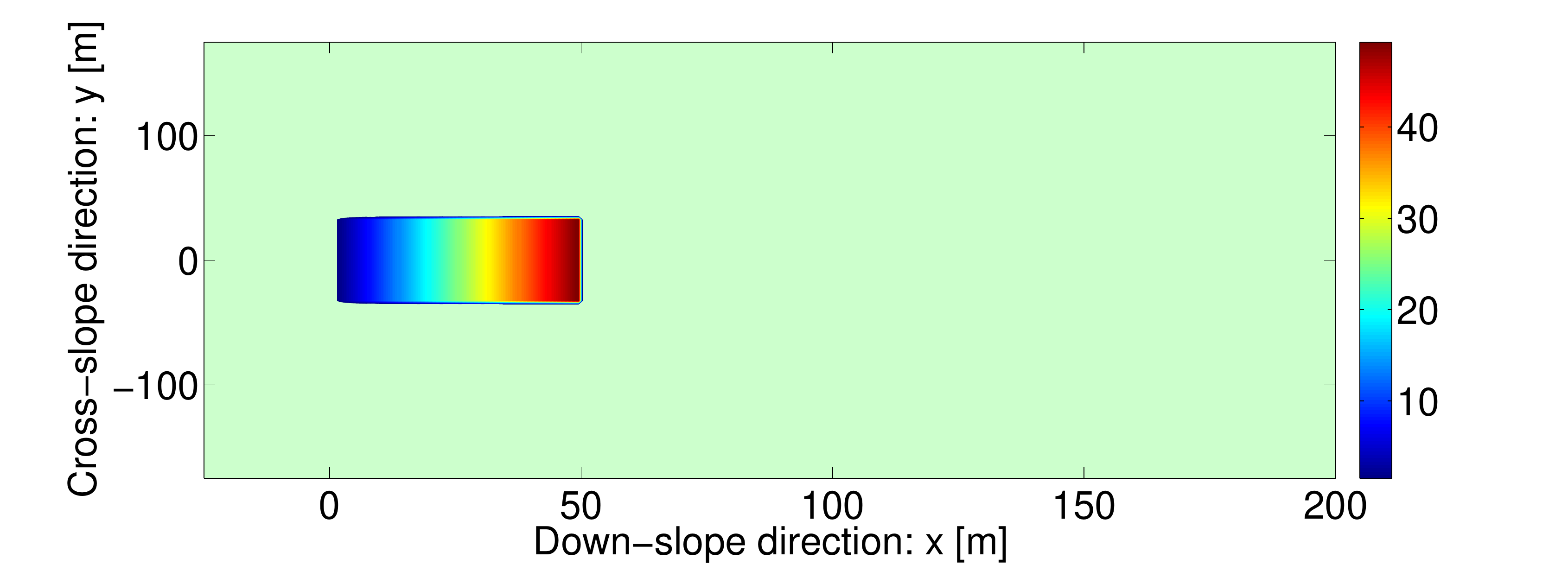}
}
\hbox{\hspace{2.5ex}
 \hspace{-11mm}
 \includegraphics[width=6.9cm]{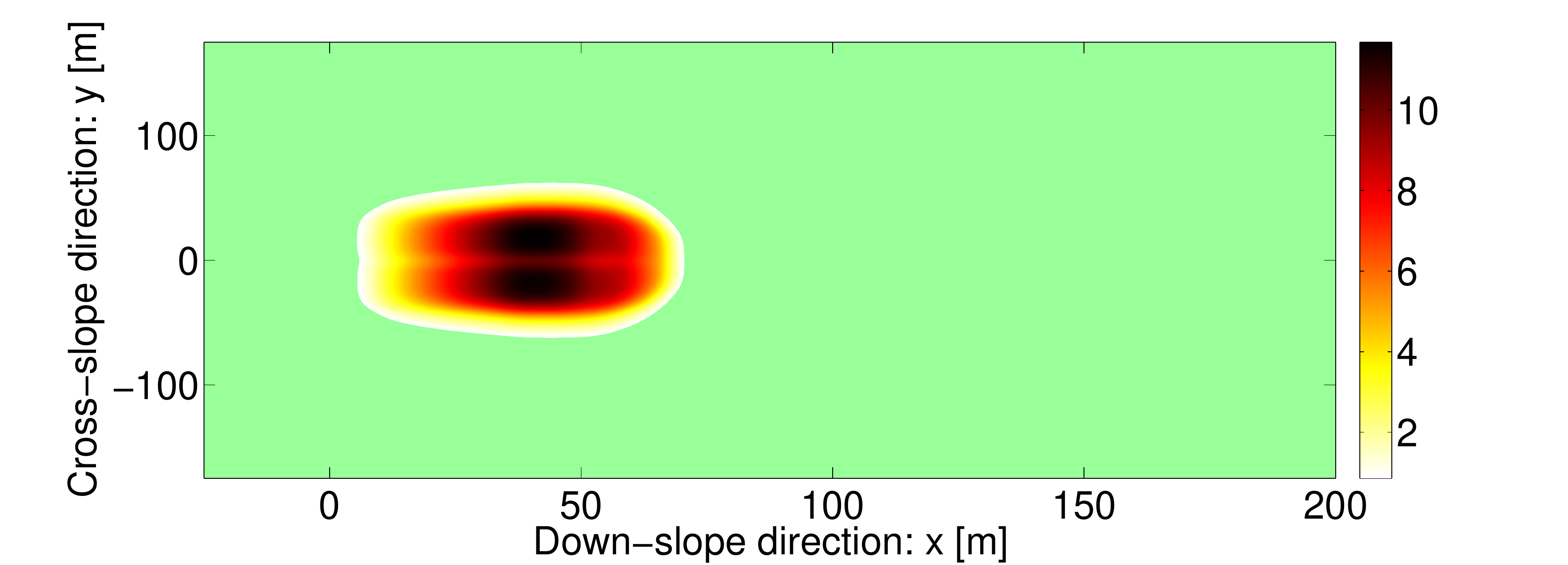}\hspace{-7mm}
 \includegraphics[width=6.9cm]{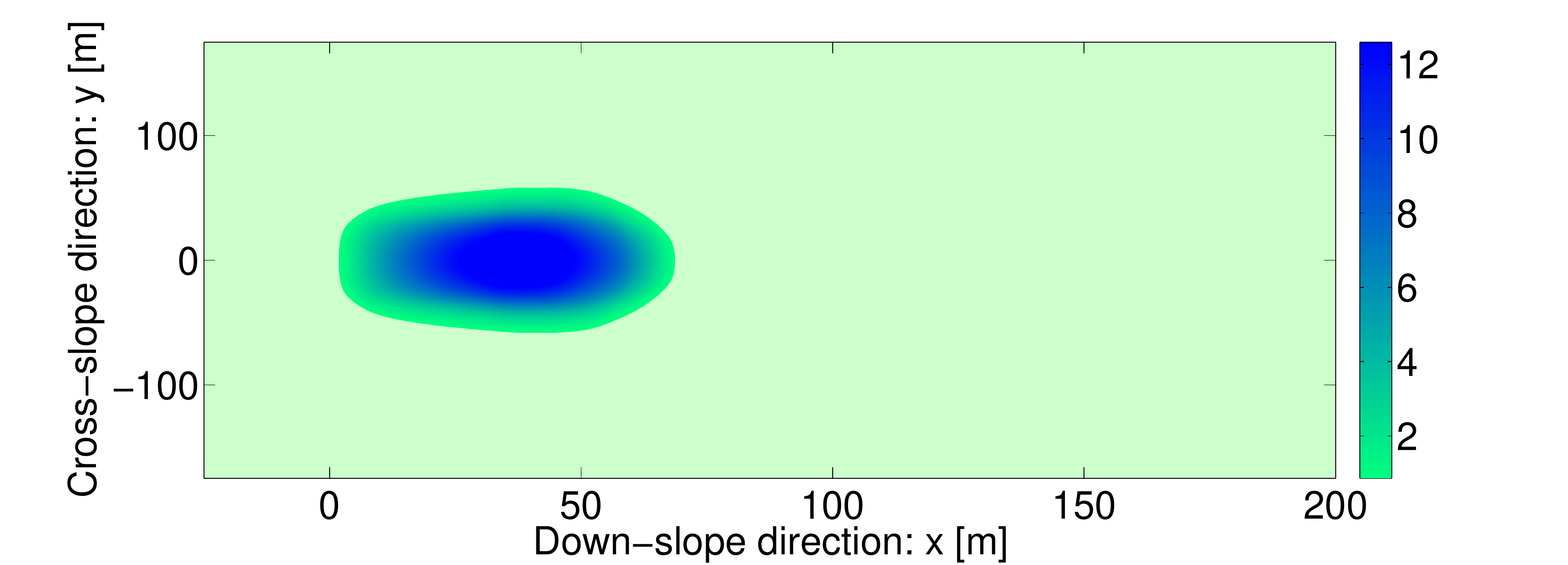}\hspace{-7mm}
 \includegraphics[width=6.9cm]{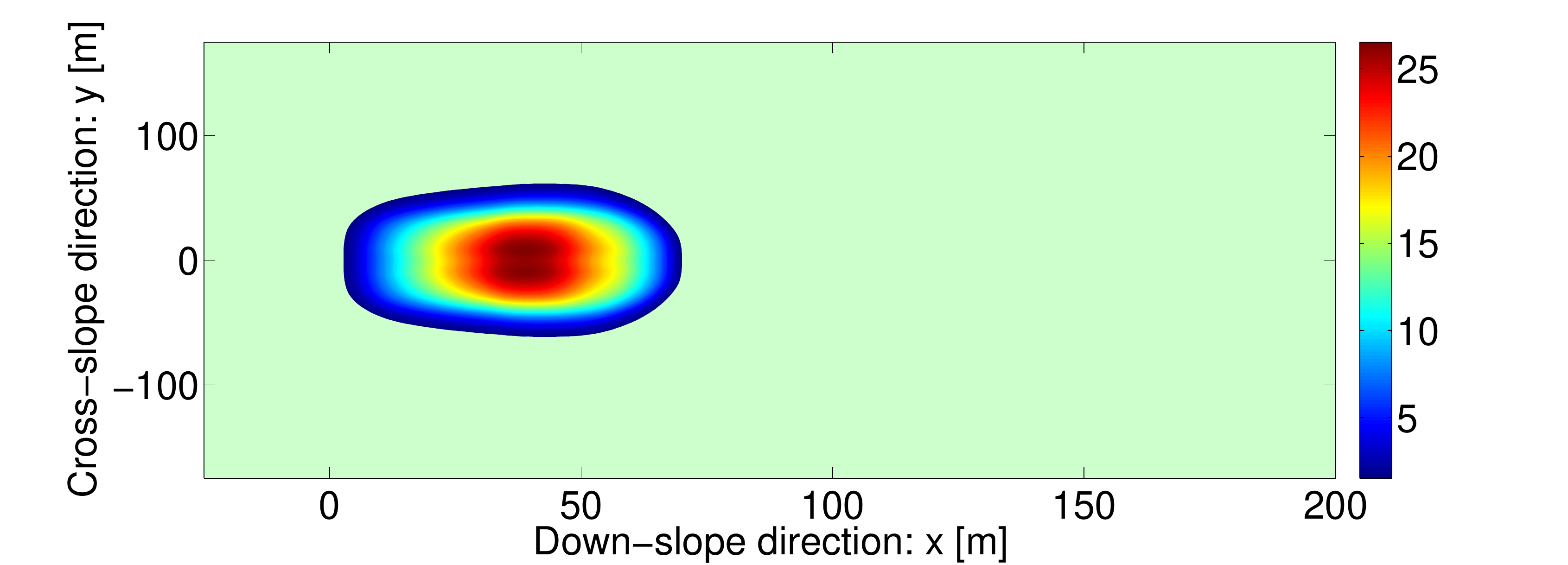}
}
\hbox{\hspace{2.5ex}
 \hspace{-11mm}
 \includegraphics[width=6.9cm]{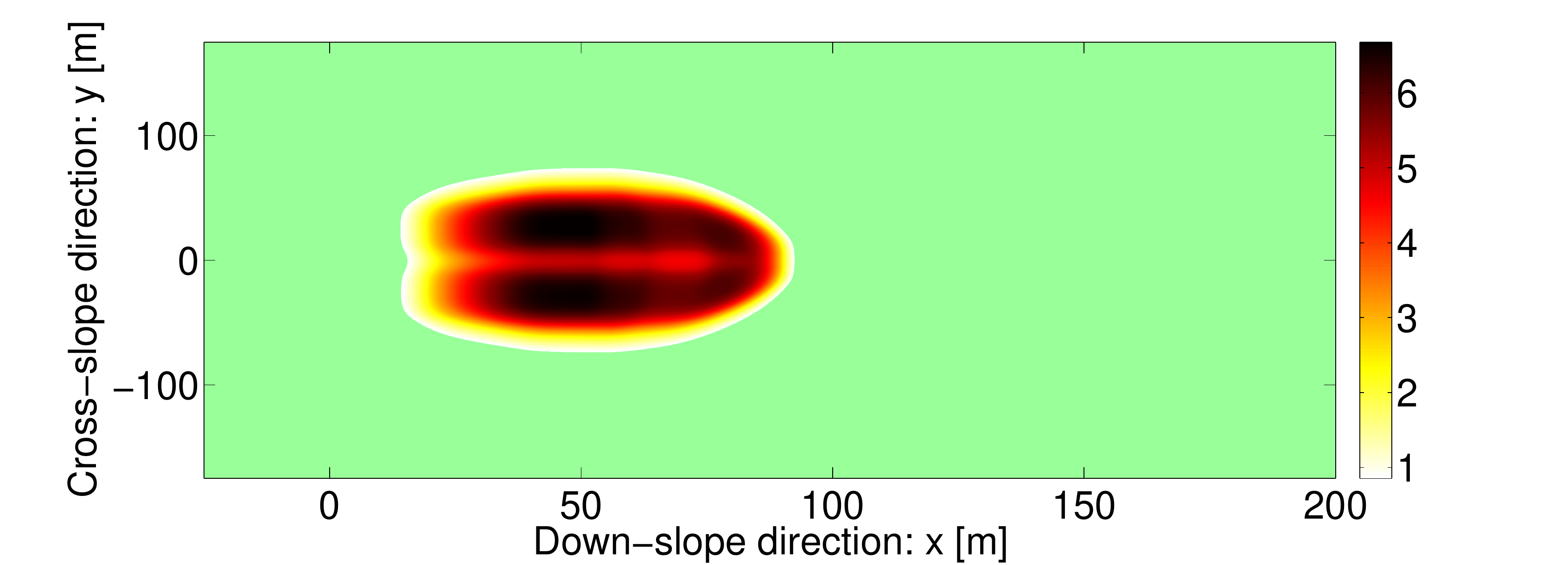}\hspace{-7mm}
 \includegraphics[width=6.9cm]{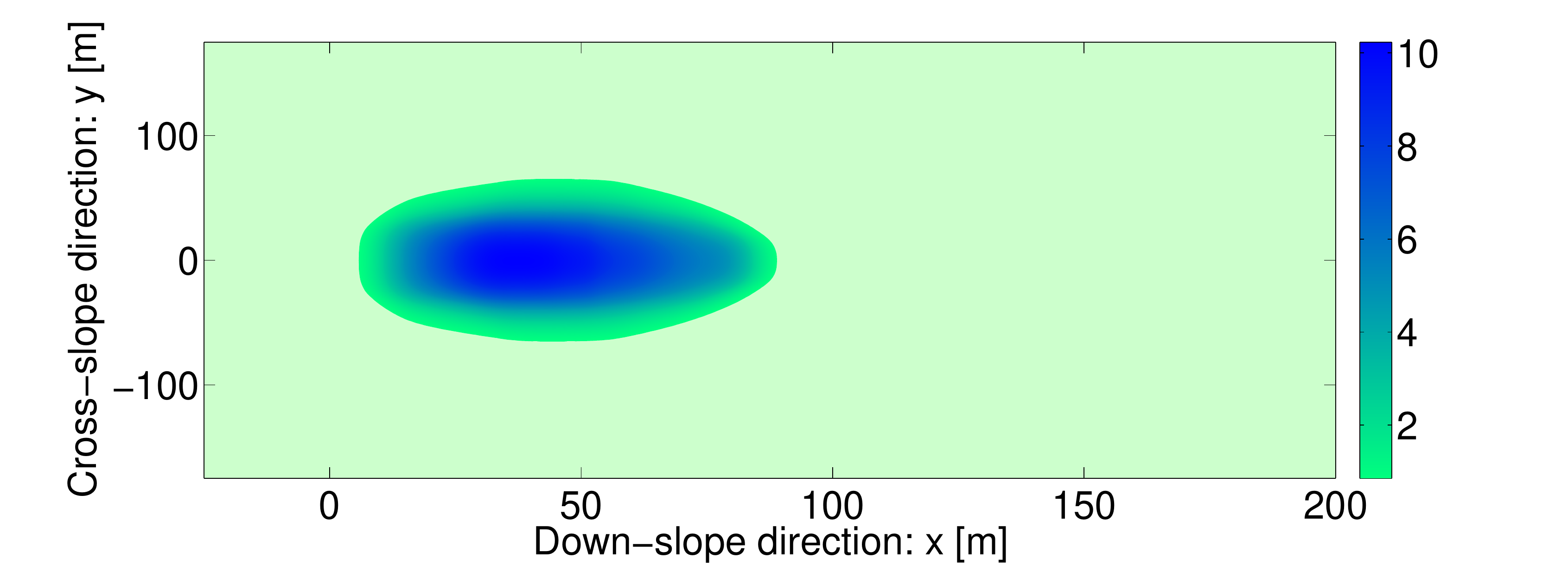}\hspace{-7mm}
 \includegraphics[width=6.9cm]{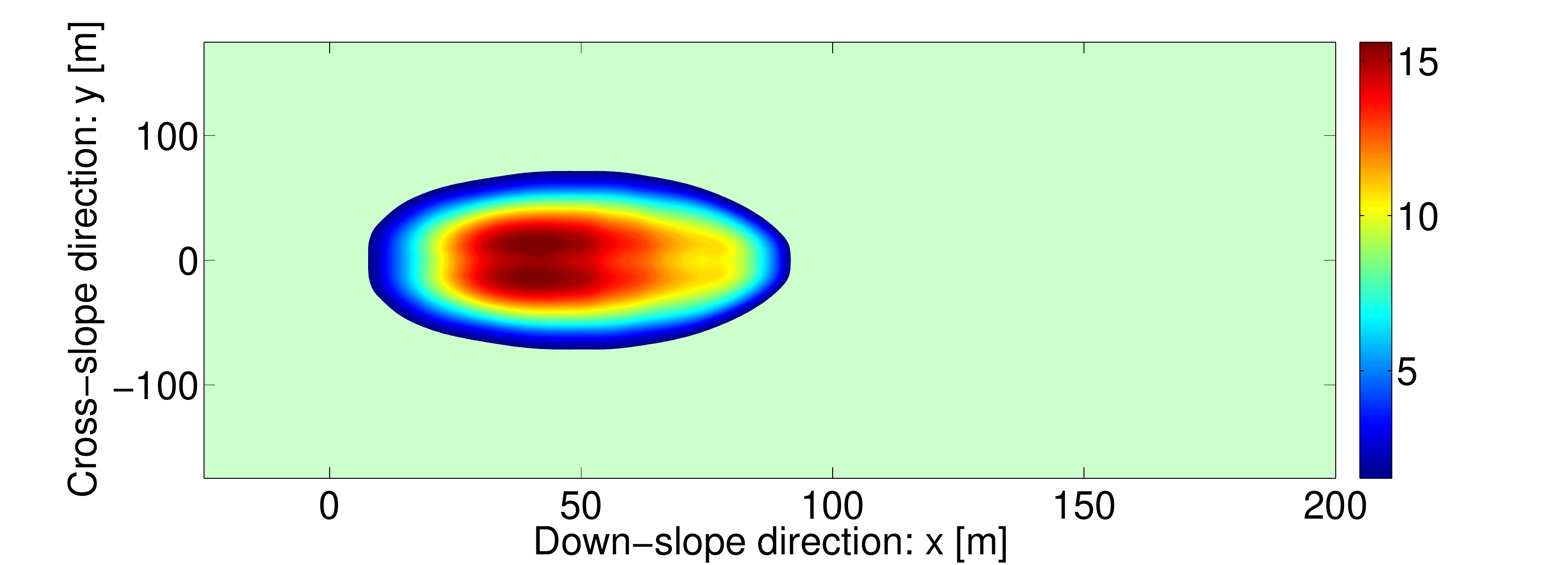}
}
\hbox{\hspace{2.5ex}
 \hspace{-11mm}
 \includegraphics[width=6.9cm]{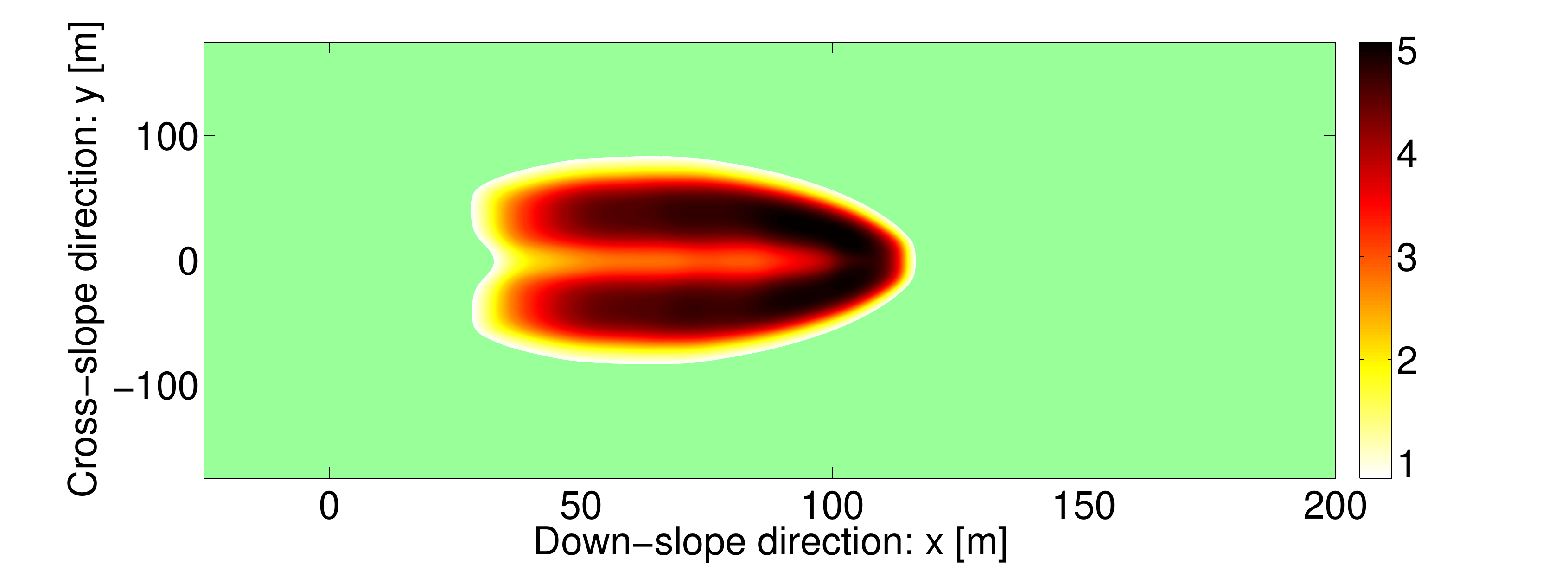}\hspace{-7mm}
 \includegraphics[width=6.9cm]{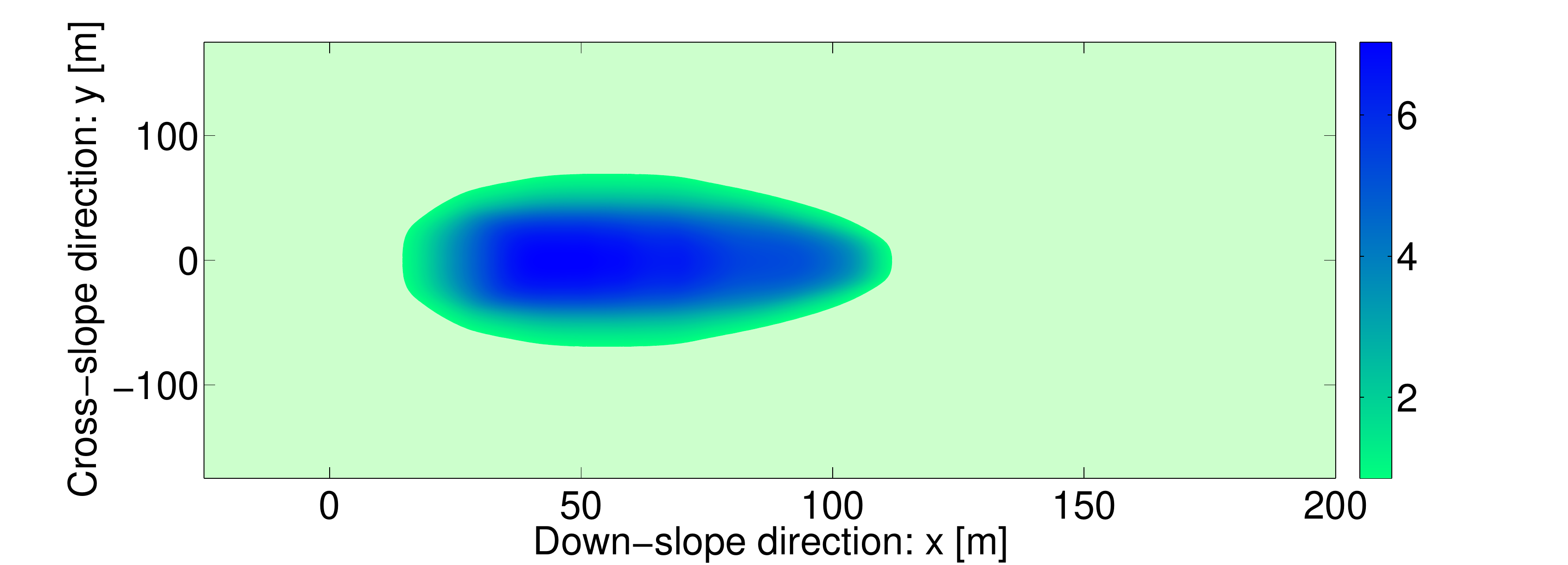}\hspace{-7mm}
 \includegraphics[width=6.9cm]{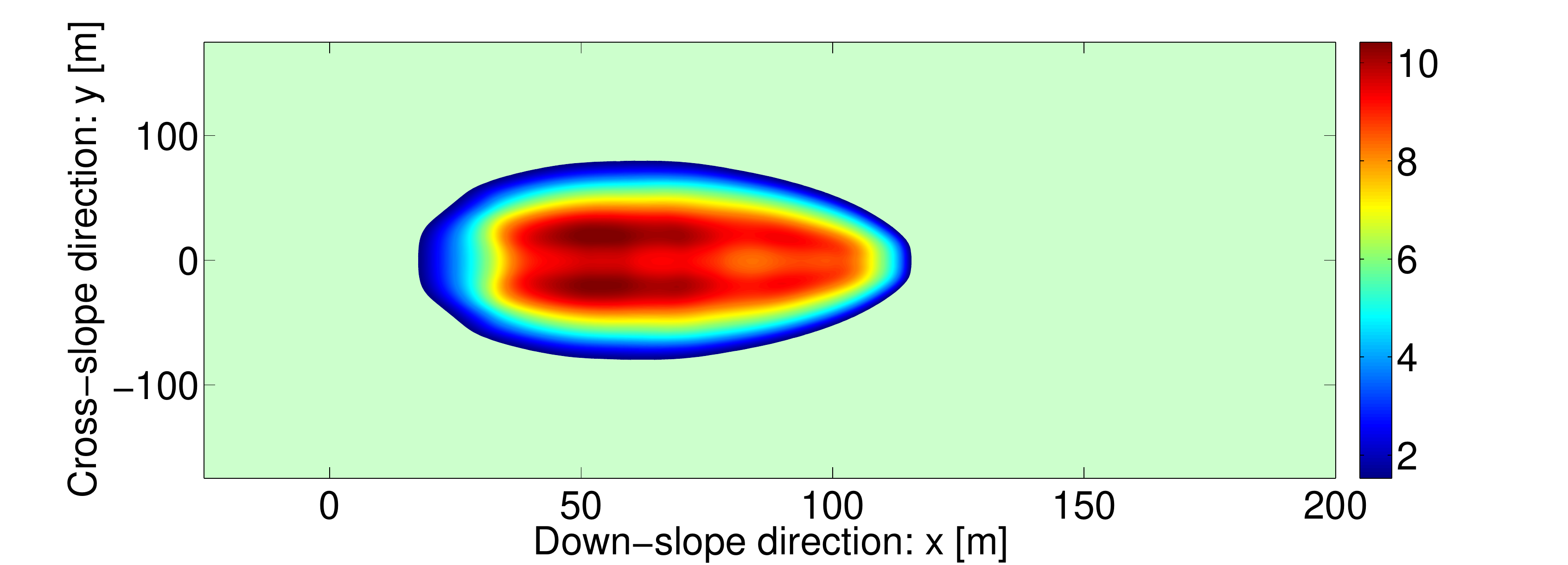}
}
\hbox{\hspace{2.5ex}
 \hspace{-11mm}
 \includegraphics[width=6.9cm]{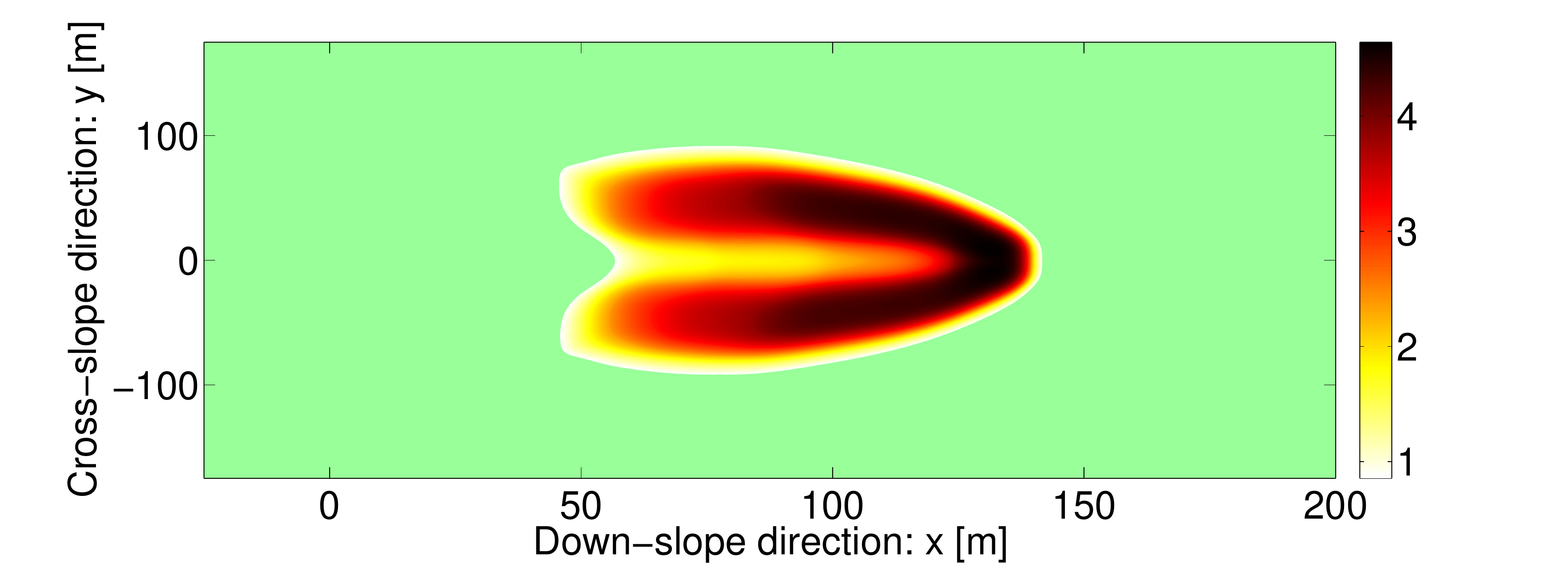}\hspace{-7mm}
 \includegraphics[width=6.9cm]{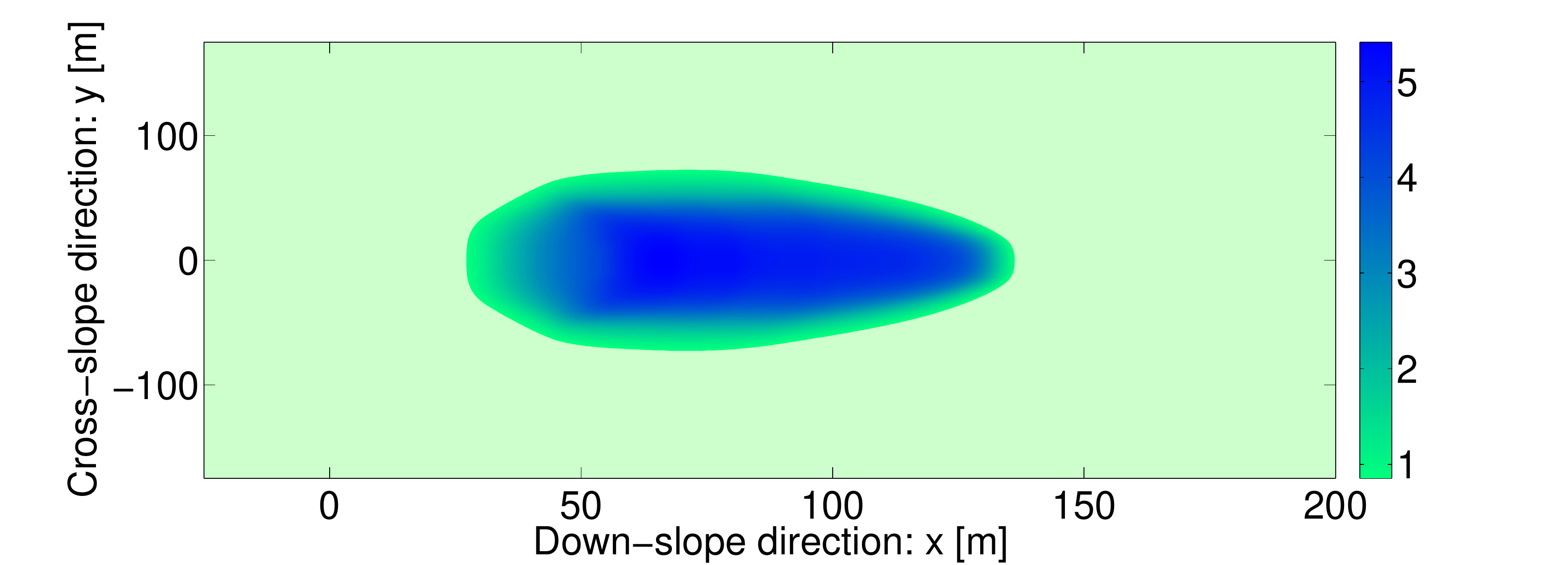}\hspace{-7mm}
 \includegraphics[width=6.9cm]{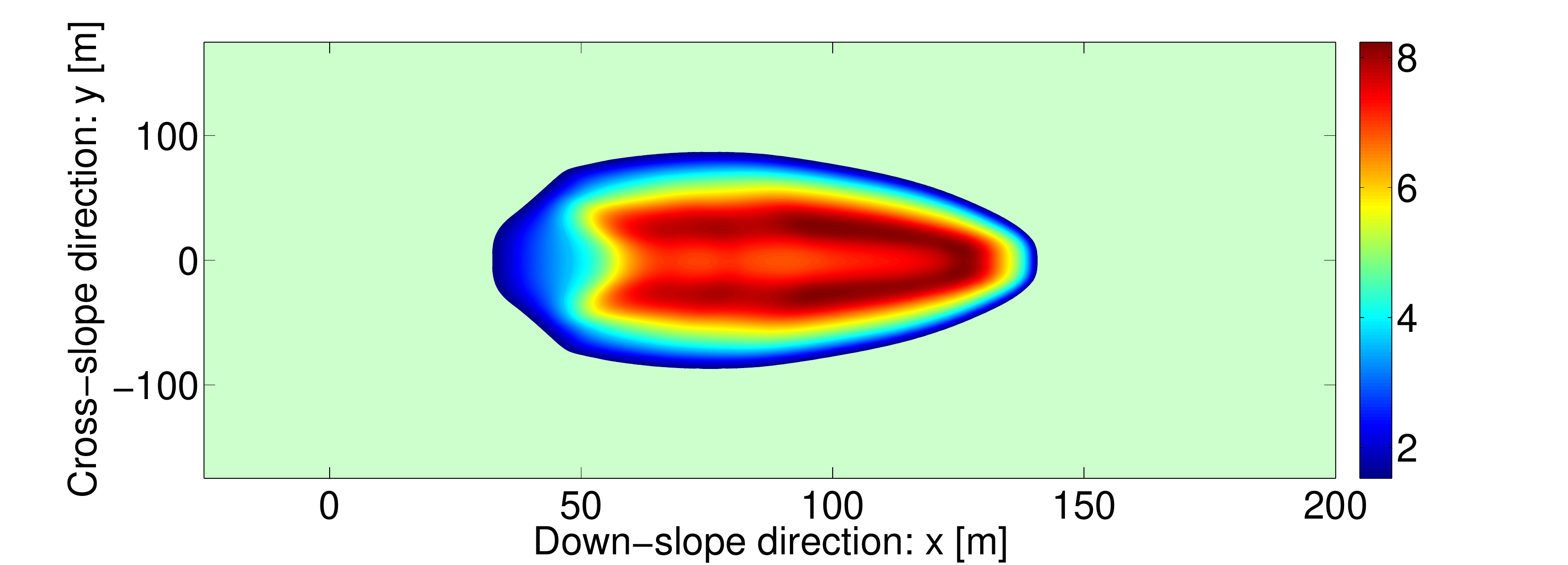}
}
\hbox{\hspace{2.5ex}
 \hspace{-11mm}
 \includegraphics[width=6.9cm]{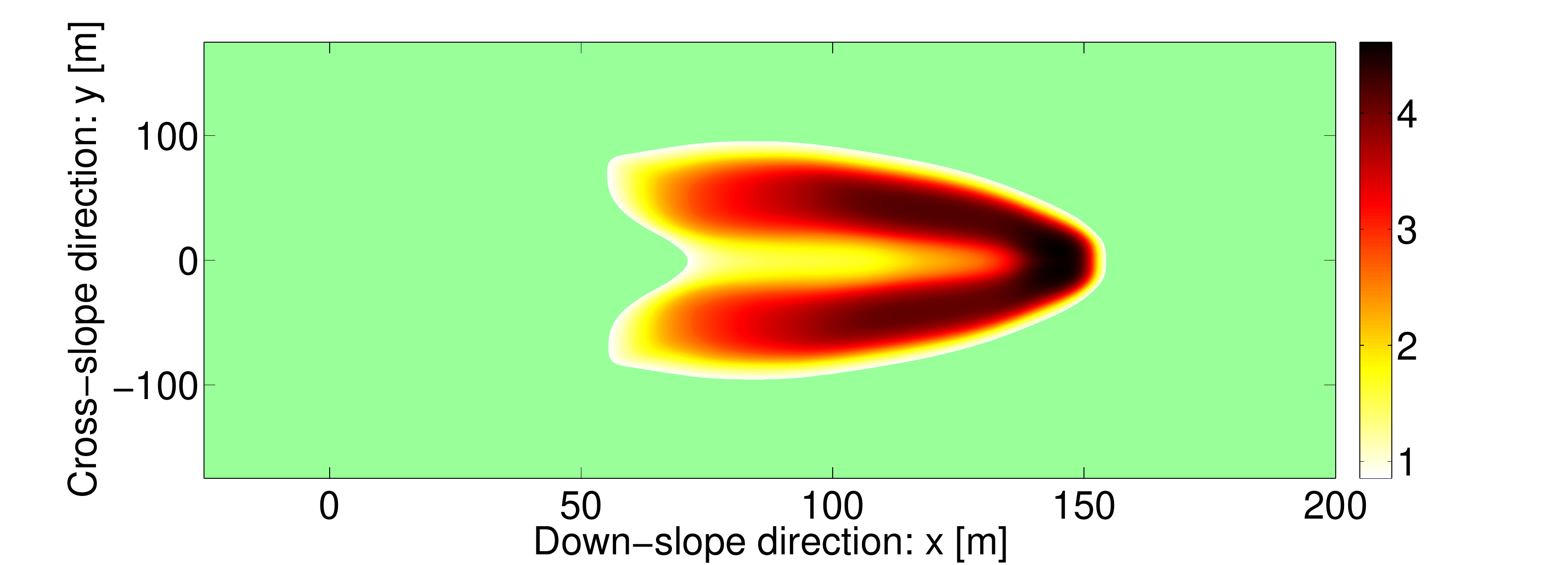}\hspace{-7mm}
 \includegraphics[width=6.9cm]{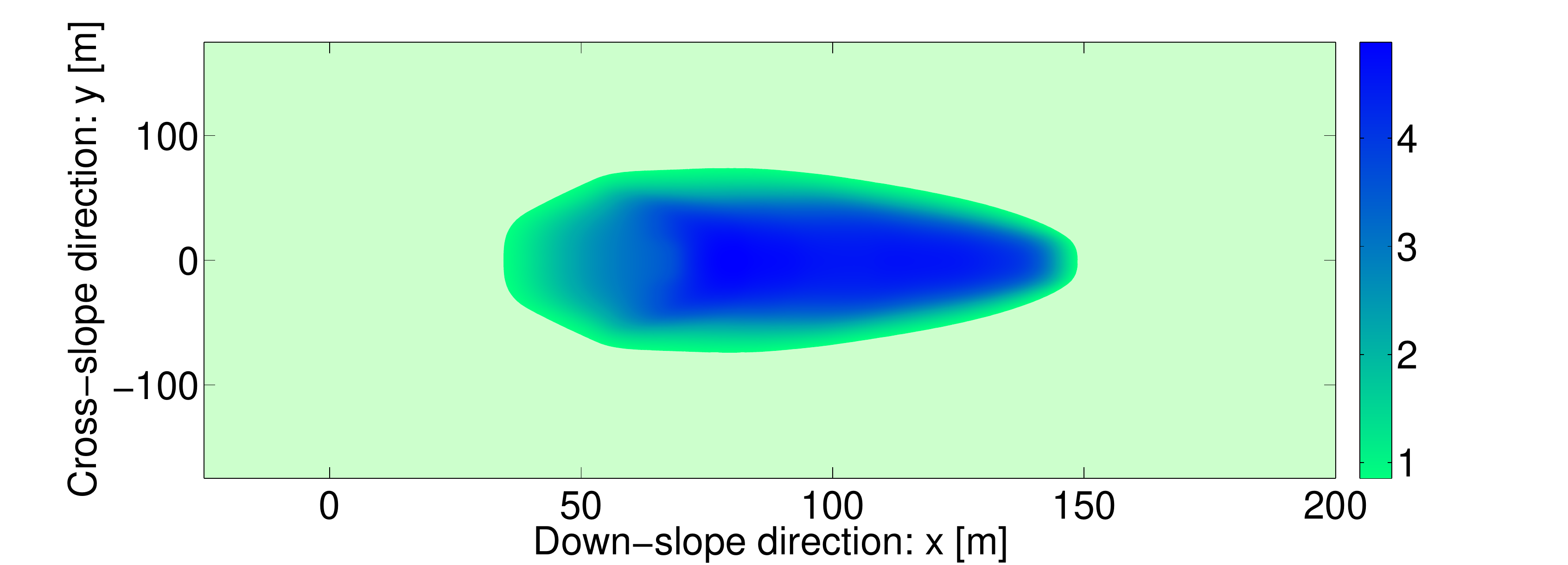}\hspace{-7mm}
 \includegraphics[width=6.9cm]{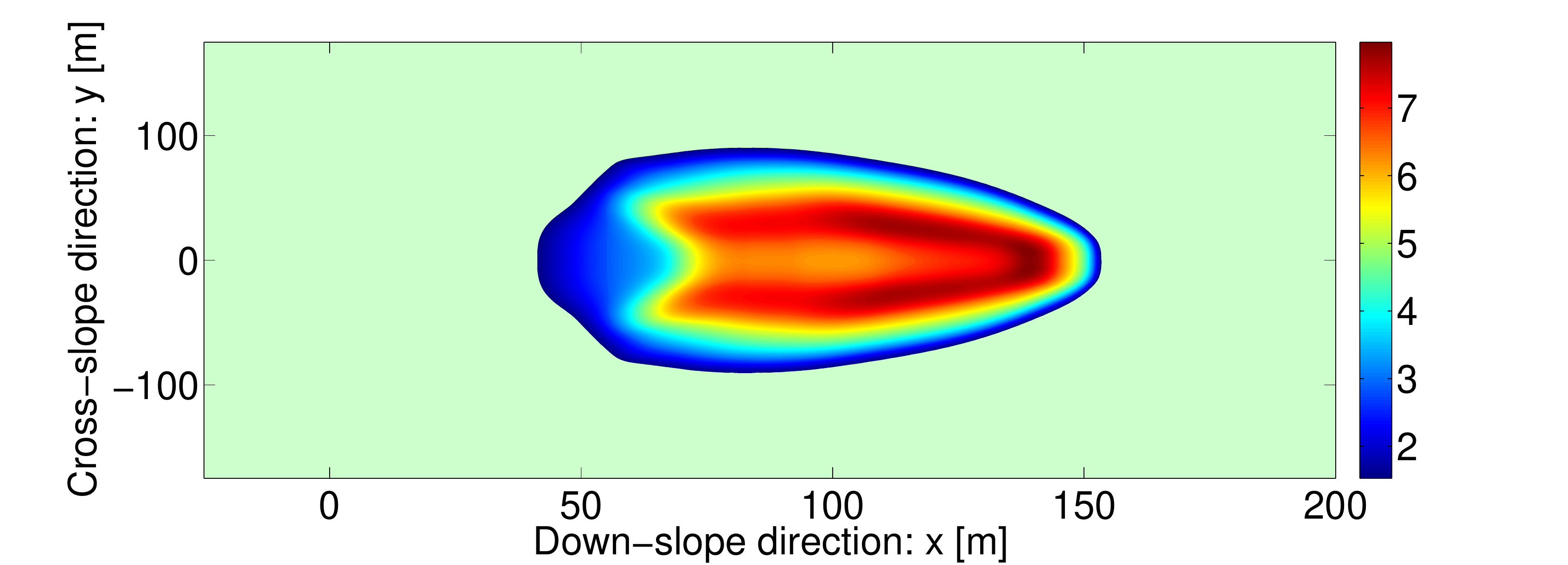}
}
\caption[]{A. Solid-, B. fluid-phase, and C. debris-mixture evolutions and their phase-separations resulting from the bi-directional separation-fluxes for solid and fluid. This clearly shows the formation and evolution of solid-rich levee and frontal-surge, and the central-and-back viscous fluid in two-phase debris flow with phase-separation as described by the novel separation-flux mechanism. Arrow indicates flow direction. Color bars are in meter.}
\label{Fig_5}
\end{center}
\vspace{-0.mm}
\begin{picture}(0,0)
\put(45,  550.){{\bf {A. Solid-Phase}}}
\put(228, 550.){{\bf {B. Fluid-Phase}}}
\put(408, 550.){{\bf {C. Total Debris}}}
\put(75,  511.){\large $\Longrightarrow$}
\put(112,  529.){\small $t = 0$\,s}
\put(112,  458.){\small $t = 1$\,s}
\put(112,  386.){\small $t = 2$\,s}
\put(112,  314.){\small $t = 3$\,s}
\put(112,  243.){\small $t = 4$\,s}
\put(112,  171.){\small $t = 4.5$\,s}
\end{picture}
\end{figure}
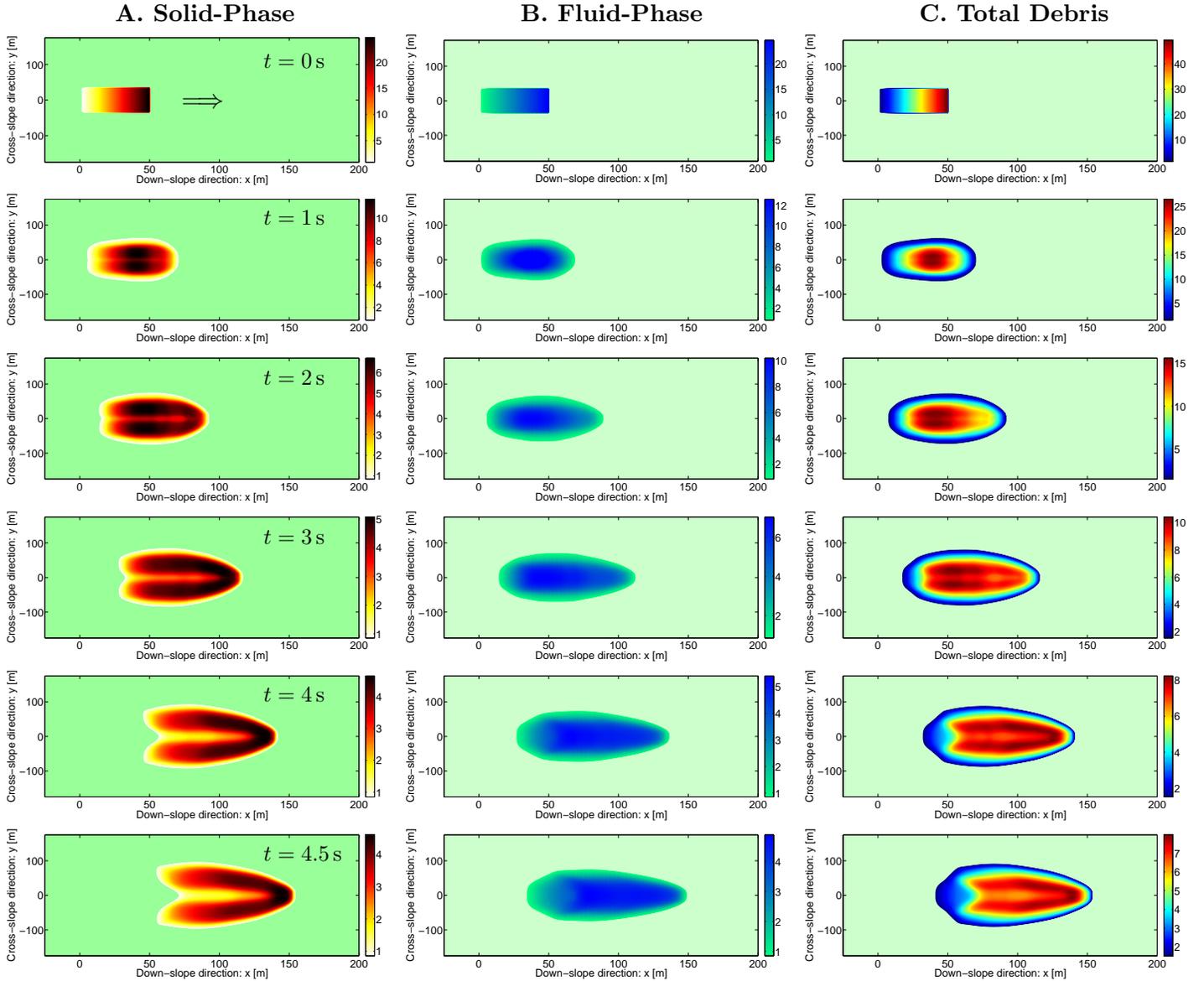
\begin{figure}[]
\begin{center}
\includegraphics[width=11cm]{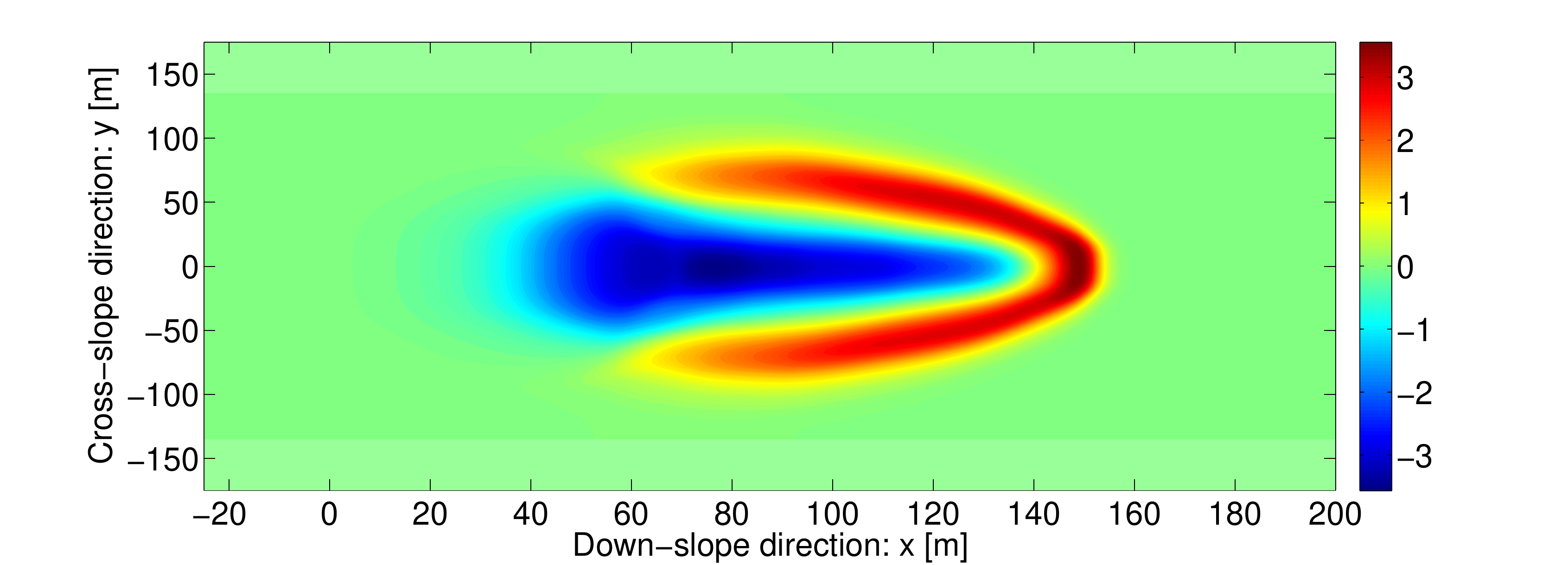}
\caption[]{Formation of levee and frontal-surge in two-phase debris flow with phase-separation as described by the novel separation-flux mechanism at time $t$ = 4.5 s. The figure is obtained by subtracting fluid-phase panel (with fluid in the central-back) from the solid-phase panel (with solid-levee and front wall) from Fig. {\ref{Fig_5}}. The figure shows the solid (in red) - fluid (blue) phase-dominance regions. Color bar is in meter.}
\label{Fig_6}
\end{center}
\end{figure}

\subsubsection{Bi-directional Fluid Phase-separation}

The middle panels B in Fig. \ref{Fig_5} display the process of fluid phase-separation which is the reverse of the solid phase-separation. Separation-fluxes for fluid bring more and more fluid to the back and central regions of the debris mass. In fact, the solid mass is transported to the lateral sides and front of the debris body. This is how these processes and solid and fluid mass evolutions (migrations) are reversed to each other. As the debris mass is released, at first slowly then rapidly, the fluid mass is accumulating in the center and back of the debris body as indicated by the amount of the fluid in panels B. The phase-vacuum, induced by the solid phase-separation, has been filled by the now separated fluid, thus, creating the viscous and particle-laden fluid (pool) in the center-and-back of the debris body. The solid concentration minimizes along the central line and decreases to the tail side while the fluid concentration is enhanced in this region. With respect to the amount of fluid, this pool is shallow in the front, side, and rear, i.e., along the frontal and lateral margins. This part of the flow is mechanically weaker as compared to the solid-rich front. So, the mechanically weakest material is in the central-back of the debris mixture. As the solid and fluid phases, and the solid and the fluid separation fluxes are coupled, the solid and the fluid structures in the left (A) and the middle (B) panels are also coupled by the complex dynamics of the solid and the fluid and their interactions.

\subsubsection{Bi-directional Phase-separations in a Debris Mixture}

Even more interesting is the dynamics of the total debris mixture as shown in the right panels C in Fig. \ref{Fig_5}. This is the combination of the solid and the fluid phases obtained by summing up the solid and fluid evolutions. Although detailed and separate knowledge of the solid and the fluid phase evolutions are very important and essential to understand the actual state of these phases and their potential domination in the total debris composition, from engineering, technical and application point of view, evolution of the total debris bulk mixture is perhaps more important than the solid only and fluid only evolutions (Kafle et al., 2016; Kattel et al., 2016). {Inundation and destructive power of the debris flow are connected to impact pressure and total energy of the mixture-debris.} For this reason, the evolution of the total debris mixture with phase separation mechanisms both for solid and fluid phases with enhanced separation-flux for solid and reduced separation-flux for fluid is presented in the right panels C (Fig. \ref{Fig_5}). Although these panels are combinations of the solid and fluid phases, they show how the separation-fluxes for the solid and fluid are interacting {with each other.}
\\[3mm]
The overall phase separation in the debris mixture is presented in the right panels C in Fig. \ref{Fig_5}. {At the early stage of the motion the solid-fluid phase-separations balance each other leading only to small changes in the total flow depth.} Nevertheless, at $t = 1$\,s, the central region shows a weak separation as that in the solid phase in the left panel. This separation intensifies at $t = 2$\,s, as the lateral levees are clearly visible, stronger in the back, and weaker in the front. For $t \ge 2$\,s, as in the solid phase, the lateral levees and frontal wall develop, evolve and consolidate. Since we have the distinct and explicit knowledge of the solid and fluid phase evolutions and their separations from each other, the right panels C, from $t = 2$\,s to $t = 4.5$\,s reveal the evolution of the solid-rich frontal surge and lateral levees with higher debris depth while the central and the back of the mixture is mainly composed of the particle-laden, viscous pool of weak (fluid) material with lower debris depth. 
    
\subsubsection{Phase-dominance} 

Besides absolute solid and fluid distributions, phase-dominance is an important aspect that appears while considering phase separation. It is defined as the difference between the solid-phase and the fluid-phase $\lb h_s - h_f\rb$, and provides the clear picture of which phase is dominating how and in which region. {The quantity $\lb h_s - h_f\rb$ could also be called the order-magnitude, since $\alpha = \lb \alpha_s - \alpha_f\rb$ is often called the order-parameter, with $\alpha\in (-1, 1)$ (Cahn and  Hilliard, 1958).} The result is shown in Fig. \ref{Fig_6} and reveals solid-rich frontal surge-head, lateral-levee and the binding-ring, while the central viscous-fluid-pool emerges behind the reinforced solid binding-ring. This is in line with observed phenomena in debris flows (Iverson, 1997; 2003; Fairfield, 2011; Battella et al., 2012; Johnson et al., 2012; Braat, 2014; De Haas et al., 2015).

\subsection{Application of the Phase-separation}
   
The evolution of the total debris mass flow dynamics in Fig. \ref{Fig_5} shows a strong solid-rich frontal surge head. Taking into account the detailed information of both, the solid and the fluid phases (panels A and B), we can now determine the strength of the material from its composition and dynamics, i.e., the accurate solid and fluid volume fractions, and their velocities. Thus, the overall strength of the debris mixture is very high in the frontal region and this strength decreases towards the tail of the debris body.
\\[3mm]
The detailed information on the evolution and distribution of the flow constituents allows to determine the dynamics and the impacts of the total debris mass. The results in Fig. \ref{Fig_5} and Fig. \ref{Fig_6} indicate that the overall strength of the debris mixture is very high in the frontal and lateral regions and decreases in the back side and the central part of the debris body. Proper knowledge of the dynamics of the phase migration (or, separation) is important, because due to its material composition (e.g., with less bulk density and viscous dissipation) the fluid-rich surge is mechanically much weaker and dynamically less destructive compared to the solid-rich surges in the frontal head. This information is important to accurately design the defense and impact dissipation structures in the debris prone regions.

\section{Summary}

The phase-separation between solid and fluid as a two-phase mass moves down slope is an often observed phenomenon and is a great challenge for scientists and engineers. To address this issue, we proposed a fundamentally new separation-flux mechanism capable of resolving the strong phase-separation in geophysical mass flows such as avalanches or debris flows. This is achieved by extending the general two-phase debris flow model (Pudasaini, 2012) with a separation-flux mechanism. The novel separation-flux model incorporates several dominant physical and mechanical aspects that result in strong phase-separation. The simulation results show that the new flux separation mechanism adequately describes and controls the dynamically evolving phase-separation and levee formation in two-phase, geometrically three-dimensional debris flows. As the separation mechanism influences the flow dynamics, solid particles are brought to the flow front and the sides. This results in a solid-rich and mechanically strong frontal surge head, and the lateral levees followed by a weaker flow body and a viscous fluid dominated tail. These phase-separation phenomena are revealed here for the first time in two-phase debris flow modeling and simulations. These simulations are in line with the field observations and laboratory experiments of debris flows.
\\[3mm]
The in-depth knowledge of the local structural and compositional evolution of the debris mixture together with the explicit picture of the solid and the fluid phases is very important for the proper understanding of the complex debris flow process. The process understanding is required to investigate the mechanical, dynamical, depositional and morphological aspects of the flow that play a vital role in structural engineering of defense structures and developing proper mitigation plans and hazard assessments in debris flow prone regions. Not taking into account these effects may result adversely as those structures cannot withstand the impact pressures exerted by the front that has much larger destructive power than that estimated in a classical fashion by single phase models, implying an uniform distribution of the composite material through the entire debris body. For typical debris flows, the solid density is about a factor three of the fluid material, and factor more than two of the mixture. Thus, just taking into account the material composition difference (neglecting the difference in the velocity evolution), impact pressure estimates of uniform mixture concepts may appear a factor two too low. This is of great importance, considering structural design of infrastructure in potentially endangered areas and justifies the application of the present model. 
\\[3mm]
{\large \bf Acknowledgements:}  This work has been conducted as part of the international cooperation projects: ``Development of a GIS-based Open Source Simulation Tool for Modelling General Avalanche and Debris Flows over Natural Topography (avaflow)'' supported by the German Research Foundation (DFG, project number PU 386/3-1) and the Austrian Science Fund (FWF, project number I 1600-N30).

\end{document}